\documentclass{aa}  

\usepackage{graphicx}
\usepackage{txfonts}
\bibpunct{(}{)}{;}{a}{}{,} 

\newcommand{\upar}{u_{\parallel}}

\newcommand{\ras}{\textsc{rascas}}
\newcommand{\vth}{\mathrm{v}_{\rm th}}
\newcommand{\lya}{Lyman$-\alpha$}
\newcommand{\toto}{\tau_{\ion{H}{i}} }
\usepackage{color}
\definecolor{v}{rgb}{0.67,0.2,0.5529}

\newcommand{\ram}{\textsc{ramses}}
\newcommand{\ramrt}{\textsc{ramses-rt}}
\newcommand{\hi}{\ion{H}{i}}
\usepackage{subcaption}

\begin{document}

   \title{RASCAS: RAdiation SCattering in Astrophysical Simulations}
   \titlerunning{RASCAS}
   \author{L. Michel-Dansac
          \inst{1}\fnmsep\thanks{\email{leo.michel-dansac@univ-lyon1.fr}}, 
          J. Blaizot\inst{1}, T. Garel\inst{1,2},
          A. Verhamme\inst{2,1}, 
          T. Kimm\inst{3}, 
          M. Trebitsch \inst{4}
          }

   \institute{Univ Lyon, Univ Lyon1, Ens de Lyon, CNRS, Centre de
     Recherche Astrophysique de Lyon UMR5574, F-69230,
     Saint-Genis-Laval, France
     \and
     Observatoire de Gen\`eve, Universit\'e de Gen\`eve, 51 Ch. des
     Maillettes, 1290 Versoix, Switzerland
     \and
     Department of Astronomy, Yonsei University, 50 Yonsei-ro,
     Seodaemun-gu, Seoul 03722, Republic of Korea
     \and 
     Sorbonne Universit\'e, CNRS, UMR 7095,
     Institut d’Astrophysique de Paris, 98 bis bd Arago, F-75014
     Paris, France 
   }

   \date{Received September 15, 1996; accepted March 16, 1997}

 
  \abstract
  {Resonant lines are powerful probes of the interstellar and
    circumgalactic medium of galaxies. Their transfer in  gas being
    a complex process, the interpretation of their observational
    signatures, either in absorption or in emission, is often not
    straightforward. Numerical radiative transfer simulations are
    needed to accurately describe the travel of resonant line photons
     in real and in frequency space, and to produce realistic mock
    observations.
  }
  {This paper introduces \ras{}, a new public 3D radiative transfer
    code developed to perform the propagation of any resonant line in
    numerical simulations of astrophysical objects. \ras{} was  
    designed to be easily customisable and to process simulations of
    arbitrarily large sizes on large supercomputers.  
  }
  {\ras{} performs radiative transfer on an adaptive mesh with an
    octree structure using the Monte Carlo technique. \ras{} features
    full MPI parallelisation, domain decomposition, adaptive
    load-balancing, and a standard peeling algorithm to construct mock
    observations. The radiative transport of resonant line photons
    through different mixes of species (e.g. \ion{H}{i}, \ion{Si}{ii},
    \ion{Mg}{ii}, \ion{Fe}{ii}), including their interaction with
    dust, is implemented in a modular fashion to allow new transitions
    to be easily added to the code.  }
  {\ras{} is very accurate and efficient. It shows perfect scaling  up to a minimum of a thousand cores.  It has been fully tested against
    radiative transfer problems with analytic solutions and against
    various test cases proposed in the literature.
    Although it was designed to describe accurately the many
    scatterings of line photons, \ras{} may also be used to propagate
    photons at any wavelength (e.g. stellar continuum or fluorescent
    lines), or to cast millions of rays to integrate the optical
    depths of ionising photons, making it highly versatile.}
   {}


   \maketitle
%

\section{Introduction}

Resonant lines are very powerful tracers of the interstellar medium
(ISM) and the circumgalactic medium (CGM) of galaxies:
their spectral and spatial distributions imprint the kinematics and
the geometry of the gas in which they scatter. The \lya{} line of
hydrogen is the brightest resonant line, extensively used to observe
statistical samples of galaxies \citep[e.g.][]{ouchi_subaru_2008,
  ouchi_2010, sobral_calymha_2017, drake_muse_2017,
  hashimoto_muse_2017, itoh_chorus_2018, herenz_muse_2018}, from the
local Universe \citep{hayes_lars_2013, henry_ly_2015, yang_ly_2017,
  verhamme_lyman_2017} to the highest redshifts
\citep[e.g.][]{zitrin_lyman_2015, stark_ly_2017}.

Other lines,  now also commonly detected in the spectra of galaxies,
contain the same richness of information as \lya\ does. In the
near-ultraviolet domain the \ion{Mg}{ii} $\lambda\lambda2796, 2804$
resonant doublet is one of the brightest. At low redshift it has been
compared to \lya\ for a small sample of Green Pea galaxies
\citep{henry_close_2018}. At intermediate redshift it has been
recently observed in MUSE-selected galaxies \citep{finley_muse_2017,
  feltre_muse_2018}, and has been compared to \ion{Fe}{ii}
\citep[$\lambda2365, \lambda2396, \lambda2612,
\lambda2626$,][]{finley_muse_2017}.  In addition, the first
\ion{Fe}{ii}$^{\star}$ extended emission has been reported
\citep{finley_galactic_2017}.

In the far-ultraviolet domain, \ion{C}{ii} $\lambda 1335$ and several
\ion{Si}{ii} lines
($\lambda 1190, \lambda1193, \lambda1194, \lambda1197$) are commonly
detected in the spectra of galaxies \citep{shapley_rest-frame_2003,
  zhu_near-ultraviolet_2015, trainor_2015, rivera-thorsen_2015,
  chisholm_2017}. Higher energy transitions are also resonant, such as
\ion{Si}{iv} $\lambda\lambda1394,1403$ and \ion{C}{iv}
$\lambda\lambda1548,1551$, probing a hotter phase of the ISM and CGM of
galaxies. With the advent of sensitive instruments such as the MUSE
Integral Field Spectrograph on the Very Large Telescope \citep[VLT,][]{bacon_muse_2010},
resonant line observations hold great potential in revealing the
chemical enrichment of the intergalactic medium (IGM) by galaxies, the
kinematics of the CGM, and the interplay between the infall and outflow
of gas around galaxies. However, resonant lines are not
straightforward to interpret. Realistic radiative transfer simulations
are needed to retrieve all information from the data.

Numerical simulations have become an indispensable tool to understand
galaxy formation. The continuous progress both in numerical simulation
algorithms and in the supercomputer platforms themselves is such that
we are today able to compute simulations of galaxies which embrace the
full complexity of their large-scale cosmological environment while
resolving the details of the ISM, and to include increasing levels of physics, such as the coupling of ionising radiation or
magnetic fields with gas \citep[e.g.][]{rosdahl_sphinx_2018,
  martin_three-phase_2018}.

In order to assess the success of a simulation and the validity of the
many assumptions that went into it, it is necessary to compare the results
with observations. This is generally not an obvious thing to do, as
simulation predictions are physical properties (e.g. density,
temperature, velocities), while observations collect photons in
various forms (spectra, images, cubes) which originate from different
processes (e.g. stellar emission, nebular emission, fluorescence) and
have possibly been altered on their way to the telescope
(e.g. scattering,
absorption by dust, IGM transmission, instrumental effects). The
maturity and high resolution (less than a few 10pc) of the current
state-of-the-art simulations offer the opportunity to produce mock
observations that account for these processes with the necessary
accuracy. In turn, provided the simulations represent the Universe
with a high enough degree of realism, these mock observations are
extremely useful for the interpretation of observations or to
test and/or predict observational diagnostics. First, they are necessary to
interpret complex observations within a robust theoretical
framework. An example here is the interpretation of diffuse \lya{}
emission seen in deep narrow-band stacks
\citep[e.g.][]{SteidelEtal11,MomoseEtal14} or in VLT/MUSE data-cubes
\citep{WisotzkiEtal16, LeclercqEtal17} in relation with the multi-band
photometry from \textit{Hubble} Space Telescope (HST). The interpretation of these phenomena requires
self-consistent multi-wavelength mocks that include the full
radiative transfer (RT)  effects due to \lya{} scattering and
dust extinction. Another example is the very difficult interpretation
of absorption lines in galaxy spectra because  observations cannot tell us
whether they are due to the ISM or CGM. Second,
mock observations are very useful to prepare and optimise
observational strategies for forthcoming surveys. Similarly,
mocks are often used before the instruments are operational as example
scenes which help validate data reduction and data analysis
pipelines. Third, mock observations may be used to validate the
accuracy with which physical properties (e.g. stellar mass, star
formation rate) are inferred from observations, since these are
known quantities from the simulations.

This  paper presents a new 3D RT code, RAdiation SCattering in Astrophysical
Simulations (\ras{}), which was designed to
construct accurate multi-wavelength mock observations (spectra,
images, or datacubes) from high-resolution simulations.  \ras{} deploys
a general two-step methodology \citep[e.g.][]{hummels_trident:_2017, BarrowEtal17}. The first step
consists in extracting all relevant information from the simulation
outputs: (1) the information concerning the medium through which light
will propagate, and (2) the information concerning the sources of
radiation. Point (1), for example, determines the number density of
\ion{H}{i} atoms, their thermal velocity dispersion and bulk velocity,
and the dust density, everywhere in a chosen volume. Fully coupled
radiation-hydrodynamic simulations would naturally provide the
information about ionised states, but retrieving this  information in
pure hydrodynamic simulations may be tricky. In such case, it is necessary
to process the simulation outputs with additional software and models,
typically with \textsc{cloudy} \citep{FerlandEtal13}, as in
\citet{hummels_trident:_2017} and \citet{BarrowEtal17}, or with
independent codes which solve for H and He ionisation by propagating
ionising radiation in post-processing \citep[e.g.][]{LiEtal08,
  yajima_art2:_2012}. Point (2) is also involved to a greater or lesser extent,
depending on the sources. Computing the continuum emission from star
particles is relatively straightforward, using spectro-photometric
models of stellar populations \citep[e.g.][]{bruzual_stellar_2003,
  eldridge_binary_2017}. However, computing the emission lines from
the gas (e.g. in the \lya{} line or in other nebular lines) again requires
 a detailed knowledge of the ionisation and thermal state of the
emitting species. This may be provided by the simulation code, as is
the case for H and He with \textsc{ramses-rt}
\citep{rosdahl_ramses-rt:_2013}, or it may be necessary to post-process the
simulation to estimate the emissivities of the gas. This first step is
very simulation- and model-dependent, and \ras{} chooses to encapsulate
it in a simulation-plugin module and to implement two stand-alone
pre-processing codes which generate  an adaptive mesh with all the
needed physical information about the gaseous medium, and the initial
conditions for light emission in the form of lists of photon
packets. These datasets, which could easily be generated from other
simulations with any post-processing code, serve as inputs to \ras{}
to perform the radiative transfer computation.

The core of \ras{}, is then the second step, which is passive
radiative transfer, passive in the sense that radiation does not
affect the properties of the gas. Here \ras{} inherits directly from
the code \textsc{McLya} \citep{verhamme_3d_2006, verhamme_lyman-_2012,
  trebitsch_lyman-alpha_2016} and joins the now long tradition of
Monte Carlo codes which compute the resonant scattering of \lya{}
photons through simulations \citep{cantalupo_fluorescent_2005,
  tasitsiomi_ly_2006, semelin_lyman-alpha_2007, laursen_ly_2009-1,
  PierleoniMaselliCiardi09, kollmeier_ly_2010, ZhengEtal10,
  laursen_intergalactic_2011, yajima_art2:_2012, behrens_effects_2013,
  smith_lyman_2015, GronkeBird17, abe_seurat:_2018} or through
idealised configurations \citep{ahn_ly_2001, zheng_monte_2002,
  dijkstra_ly_2006, hansen_lyman_2006, BarnesHaehnelt09,
  Forero-RomeroEtal11, OrsiLaceyBaugh12, lake_diffuse_2015,
  gronke_ly$alpha$_2016,eide_unlocking_2018}.  The novelty of \ras{}
compared to these \lya{} RT codes is a combination of the following
qualities: (1) \ras{} is a public code and, being the latest
implementation, incorporates a selection of state-of-the-art
algorithms presented in other codes; (2) \ras{} is designed in a very
modular way and can be used for \lya{} RT, but also for computing
radiative transfer of any other resonant (or not) line and continuum
radiation (with arbitrary spectral shapes); (3) \ras{} is massively
parallel and implements a domain decomposition which allows it to run
with a negligible memory footprint on many cores, thus making it a
perfect tool to process extremely large simulations on very large
supercomputers.

The layout of the paper follows the Monte Carlo workflow. In
Sect.~\ref{sec:sampling} we describe how we sample radiation
sources with photon packets. In Sect.~\ref{sec:cellrt} we
present our methods for solving radiative transfer in a single
simulation cell. In Sect.~\ref{sec:meshrt} we detail our
parallel implementation on adaptive grids with domain
decomposition. We then present brief illustration for possible
applications in Sect.~\ref{sec:applications} and sum up in
Sect.~\ref{sec:summary}.

\section{Emission of photon packets}
\label{sec:sampling}

In this section, we describe how we generate photon packets from
various sources of radiation. In doing so, we  sample the
spectral and angular distributions which characterise each source
with a number of photon packets that relates to the luminosities of
the sources. In the current implementation, all sources considered (star
particles or gas cells) are isotropic, and photon packets are emitted
with random directions. Anisotropic sources may be useful, for example
to describe an active galactic nucleus or to set up idealised experiments with specific
illuminations (e.g. plane-parallel, beam). The framework of
\ras{} makes this trivial.  \ras{} also gives an equal weight to all
photon packets so that they are cast from different sources with a
probability $P = \dot{N}_\gamma/\dot{N}_\gamma^{\rm tot}$, where
$\dot{N}_\gamma$ is the number of (real) photons emitted by a source
per unit time and $\dot{N}_\gamma^{\rm tot}$ the sum of $\dot{N}_\gamma$ over all
sources. We detail below how we compute these terms in different
configurations.

In the framework of \ras{}, this first step is done by a stand-alone
code which generates a simple photon packet initial condition (PPIC)
file for each experiment. This file contains a list of photon packets, each with an ID, a position, a direction, and a frequency. These
PPIC files are then fed to \ras,{} which propagates each photon packet
regardless of the sources. This very flexible approach is possible
because the radiative transfer done with \ras{} is passive (i.e. it
does not change the properties of the gas). It allows  statistics
to be added incrementally by re-generating additional PPIC files on demand, and it
makes it very easy to combine   different signals (e.g. stellar
continuum and nebular lines) run as separate experiments. The
simplicity of these PPIC files also renders trivial the generation of
any ad hoc source model.

Three stand-alone codes are currently available to generate PPIC files
for different sources. These codes, which may serve
as templates for future developments by interested users, are
described in the following subsections.

\subsection{Emission from stars}
The first stand-alone code, {\tt PhotonsFromStars}, generates photon
packets from star particles which represent stellar populations of
single ages and metallicities. Four spectral shapes are implemented:
(1) a power-law fit to the stellar continuum, (2) a tabulated version
of the stellar continuum, (3) a Gaussian emission line, and (4) a
monochromatic line. In all cases, the stellar emissivity from star
particles is computed with a 2D interpolation in age and metallicity
of tabulated spectro-photometric models such as
\citet{bruzual_stellar_2003} or \citet{eldridge_binary_2017}.

The frequencies of photon packets are defined in the rest frames of
the sources and then shifted to an external frame according to each
source's velocity. We give details below for the different spectral
shapes.

\subsubsection*{Power-law continuum} 

Here we consider a continuum flux density described by a power law
$F_\lambda = F_0 (\lambda/\lambda_0)^\beta$ between two boundaries
$\lambda_{\rm min}$ and $\lambda_{\rm max}$. Both $F_0$ and $\beta$
depend on age and metallicity and are different for each star
particle. The number of photons emitted per unit time by this
continuum is
\begin{eqnarray} \label{eq:one}
\dot{N}_\gamma=
\begin{cases}
\frac{F_0 (\lambda_{\rm max}^{\beta+2}-\lambda_{\rm min}^{\beta+2})}{hc \lambda_0^\beta(\beta+2)} & \; {\rm if} \; \beta \neq -2, \\
\frac{F_0\lambda_0^{2}}{hc } \ln(\lambda_{\rm max}/\lambda_{\rm min})
& \; {\rm if} \; \beta = -2, \\
\end{cases}
\end{eqnarray}
and each source will emit on average
$N_{\rm MC} \times \dot{N}_\gamma/\dot{N}_\gamma^{\rm tot}$ photon
packets, where $N_{\rm MC}$ is the total number of photon packets
generated. In practice, once $\dot{N}_\gamma$ is known for all
sources, we store the normalised cumulative luminosity distribution of
sources into an array $P$, so that the $i$-th element of the array,
$P_i$, has a value equal to the probability of emission by a particle
with index $\leq i$. Then, for each photon packet, we draw a
univariate $r$ between 0 and 1, and locate $r$ in the array $P$ with a
bi-section algorithm so that $P_{i-1} < r \leq P_i$. This gives us
the index $i$ of the star particle which will emit the photon packet.

When a star particle emits a photon packet, we derive a wavelength by
sampling the photon number distribution
$P(\lambda) = \lambda F_\lambda \dot{N}_\gamma / hc$. This function is
integrable analytically, so that the probability of emitting a photon
with wavelength in $[\lambda_{\rm min};\lambda]$ is written
$P(<\lambda) = (\lambda^{2+\beta} - \lambda_{\rm
  min}^{2+\beta})/(\lambda_{\rm max}^{2+\beta} - \lambda_{\rm
  min}^{2+\beta})$ for $\beta \neq -2$, and
$P(<\lambda) = \ln(\lambda/\lambda_{\rm min}) / \ln(\lambda_{\rm
  max}/\lambda_{\rm min})$ for $\beta = -2$. These two expressions can
be inverted, and we simply need to draw a univariate $r$ between 0 and
1 to compute the wavelength $\lambda$ of the photon packet as
\begin{eqnarray} \label{eq:two}
\lambda = 
\begin{cases}
\left[ \lambda_{\rm min}^{2+\beta} + r \times
  (\lambda_{\rm max}^{2+\beta} - \lambda_{\rm min}^{2+\beta})
\right]^{1/(2+\beta)} & \; {\rm if} \; \beta \neq -2, \\
\lambda_{\rm min} \times e^{r \ln(\lambda_{\rm max}/\lambda_{\rm min})}
& \; {\rm if} \; \beta = -2. \\
\end{cases}
\end{eqnarray}

\subsubsection*{Tabulated continuum from stellar populations} 
To obtain more realistic spectra including, for example, stellar atmosphere
absorption features, we follow the same strategy as above with small
modifications. We replace Eq.~\ref{eq:one} with numerical estimates
computed directly from the tabulated spectra provided by
spectro-photometric libraries, using a linear interpolation of
$F_\lambda$ in wavelength. As above, this first step allows us to
assign photon packets to sources as a function of their ages and
metallicities. Then, in order to define the wavelength of a photon
packet, we store in an array $P$ the normalised cumulative number of
photons (per unit time) emitted by its source between
$\lambda_{\rm min}$ and $\lambda_i$, where $\lambda_i$ are the values
at which the library provides $F_\lambda$ values. Drawing one
univariate allows us, with a  strategy similar to that above, to locate the
wavelength bin in which the photon packet is emitted. Once this is
known, we keep following the procedure  by drawing a second 
univariate to compute the wavelength of the photon packet using Eq.~\ref{eq:two} where the
power law is now linear and describes $F_\lambda$ in the range
$[\lambda_{i-1};\lambda_i]$.
This method produces a
linearly interpolated version of the input tabulated energy
distribution. 

\subsubsection*{Gaussian line} 
In  some experiments the aim might be  to model the emission lines from
star particles, for example as a proxy for nebular emission from
\ion{H}{ii} regions
\citep[e.g.][]{verhamme_lyman-_2012,behrens_braun_2014}. In such
cases, $\dot{N}_\gamma$ is the number of photons emitted per second in
the line, and can be derived from the number of ionising photons
emitted by each star particle as a function of its age and
metallicity. We follow here the same strategy as above, with
normalisation factors which depend on the nebular line under
consideration, and assign a source for each photon packet.

In  order for frequencies of photon packets to be distributed along a
Gaussian centred on frequency $\nu_0$ and of standard deviation
$\sigma_\nu$, we use the Box-Muller method: we draw two random numbers
$r_1$ and $r_2$ uniformly distributed between 0 and 1 and compute
$\nu =\nu_0 (1 + \sqrt{- 2\ln(r_1)} \cos(2\pi r_2) \sigma_\nu)$. We note 
that this implementation makes the approximation that photons have the
same energy across the line (in the frame of each source). This is
generally  acceptable for lines broadened by a characteristic
velocity $\mathrm{v} \ll c$, with an error in $\mathrm{v}/c$.

\subsubsection*{Monochromatic line} 
A final case of interest is monochromatic emission from star
particles. This may be used for at least two purposes: (1) a
simplification of line emission and (2) a way to obtain a quick estimate
of the continuum flux at any wavelength. In case (1) the same
procedure as Gaussian emission is followed, but the frequency of all
photon packets are the same (in the frame of each source). In (2)  the emitted number of photons per unit time is used at the chosen
wavelength $\lambda_{\rm em}$,
$\dot{N}_\gamma = \lambda_{\rm em}F_\lambda(\lambda_{\rm em}/hc)$, to
assign sources to photon packets, and again the frequencies of all
photon packets are the same.

\subsection{\lya{} emission from gas}
The second stand-alone code, {\tt LyaPhotonsFromGas}, generates photon
packets which describe \lya{} emission from the gas\footnote{This code
  can be easily customised to deal with any other emission line from
  the gas.}. This code was written to exploit \ramrt{} simulations, where the ionisation
state of hydrogen is known, so that \lya{} emission can be computed
directly from the simulation outputs. We note that there are two
contributions to \lya{} emission: recombinations and collisions. We
treat the two separately, and generate an independent PPIC file for
each.

This time the sources are  adaptive mesh refinement (AMR) cells, not
particles, and we assume that emission is homogeneous within each
cell. We follow a similar strategy to that for star particles, and first
we compute the number of \lya{} photons emitted per unit time by each
cell. For recombinations, we assume the case B conditions and,
following \citet{cantalupo_2008}, we compute the number of \lya{}
photon emitted per unit time from each cell as
\begin{equation} \label{eq:LyaCellLum} 
\dot{N}_{\gamma,{\rm rec}} = n_e n_p \epsilon^{\rm B}_{\rm Ly\alpha}(T)
\alpha_{\rm B}(T) \times (\Delta x)^3,
\end{equation}
where $n_e$ and $n_p$ are the electron and proton number densities,
predicted by our simulation as a function of the local radiation field
and read directly from \ramrt{} outputs; $\alpha_{\rm B}(T)$ is the
case B recombination coefficient; $\epsilon^{\rm B}_{\rm Ly\alpha}(T)$
is the fraction of recombinations producing \lya{} photons; and
$(\Delta x)^3$ is the volume of the cell. We evaluate
$\epsilon^{\rm B}_{\rm Ly\alpha}(T)$ with the fit from \citet[][their
Eq. 2]{cantalupo_2008}, and $\alpha_{\rm B}(T)$ with the fit from
\citet[][their appendix A]{hui_gnedin_1997}. For the collisional term,
we compute the number of \lya{} photons emitted per unit time from
each cell as
\begin{equation}
\dot{N}_{\gamma,{\rm col}} = n_e n_{\rm \ion{H}{i}} C_{\rm Ly\alpha}(T)  \times (\Delta x)^3,
\end{equation}
where $n_{\rm HI}$ is the number density of neutral hydrogen atoms,
read directly from the simulation outputs, and $C_{\rm Ly\alpha}(T)$
is the rate of collisional excitations from level $1s$ to level $2p$
which we evaluate with the fit from \citet[][their
Eq. 10]{goerdt_2010}. Once these luminosities are known for all cells,
we use the same algorithm as for star particles to assign photon
packets to cells.

Finally, for each photon packet, we generate random emission
coordinates within its emission cell, and compute an emission
wavelength in the frame of the cell following the Gaussian line method
outlined for star particles. In this case, we relate the width of the
Gaussian to the mean velocity of hydrogen atoms through
$\sigma_\nu = \nu_0 (2k_BT/m_p)^{1/2} / c$, where $k_B$ is Boltzmann's
constant and $m_H$ the mass of a hydrogen atom.

\subsection{Ad hoc source models}

The simplicity of PPIC files makes it straightforward to write these
files with any idealised configuration. A few examples are implemented
in the \ras{} distribution, which can be used as they are or extended
to different configurations.

\section{Radiative transfer in a homogeneous medium}
\label{sec:cellrt}

In this section we discuss the implementation of Monte Carlo
radiative transfer (MCRT) through a homogeneous medium
(e.g. a simulation cell). Unless mentioned otherwise, we compute
frequencies and velocities in the frame of that medium. Extending this
to the general case is simply a matter of representing a complex gas
distribution on a grid (see Sect.~\ref{sec:meshrt}) and dealing with
changes of frames from cell to cell. To the first order in
$(\mathrm{v}_g/c)$, where $\vec{\mathrm{v}}_g$ is the velocity of the
gas relative to some external frame,  the frequency of a
photon packet in the gas frame $\nu$ can be related to that in the external frame
$\nu_{\rm ext}$ with
$\nu_{\rm ext} = \nu(1 + (\vec{\mathrm{k}} \cdot \vec{\mathrm{v}_g})/c)$,
where $\vec{\mathrm{k}}$ is the propagation direction vector.

Photon packets propagate along straight lines between interactions
with matter or until they escape the computational domain towards the
observer. After emission, or after each scattering, the optical depth
to the next scattering event is drawn from an exponential distribution
using
\begin{equation}
\tau_{\rm event} = -\ln(r),
\end{equation}
where $r$ is a uniformly distributed random number. 
The photon packet is advanced in space until the optical depth $\tau$
it covered reaches $\tau_{\rm event}$, at which point the interaction
occurs. In practice, we propagate photon packets through a grid, and
they may have to cross more than one cell before reaching
$\tau_{\rm event}$. In such cases, photon packets are moved from cell
to cell, each time subtracting to $\tau_{\rm event}$ the contribution
$\tau_{\rm cell}$ across the previous cell.

The computation of $\tau$ along a ray is discussed in
Sect.~\ref{sec:tau}. Different interactions implemented in \ras{} are
then described in Sect.~\ref{sec:interactions}. After this overview, we
give details of the numerical implementations in \ras{} and test them
in Sect.~\ref{sec:mcrt}.

\subsection{Optical depths and  determining the next interaction}
\label{sec:tau}

The total optical depth $\tau_{tot}$ through a mixture of gas and dust
can be written as
\begin{equation}
\tau_{\rm tot}(\nu) = \sum_{X}^{\rm species} \sum_{lu}^{\rm transitions} \tau_{X,lu} (\nu) +
\tau_{\rm dust}(\nu), 
\end{equation}
where $\nu$ is the photon frequency in the frame of the gas (and dust)
mixture. The first term on the  right-hand side   is a sum over all transitions,
from a lower level $l$ to an upper level $u$, from all atomic or ionic
species $X$ (hereafter scatterers) present in the gas. The computation
of these resonant line optical depths is developed in
Sect.~\ref{sec:tau_line}.  The second term is the contribution of dust,
discussed in Sect.~\ref{sec:tau_dust}. We note that \ras{} does not
require any line to be present in the above sum, and it can also be used
to propagate continuum through a dusty medium. We also note that other
continuum contributions can be added to this sum, depending on
frequency and/or scatterer. For instance, the absorption in the Lyman
continuum can be taken into account by adding a term to this sum
\citep[e.g.][]{inoue_monte_2008}.

\subsubsection{Resonant line optical depth} \label{sec:tau_line} 

Along a path of length $\ell$ through gas at temperature $T$, the
optical depth due to the transition from a lower level $l$ to an upper
level $u$ of scatterer $X$ may be written as a function of frequency
$\nu$ in the gas frame as
\begin{equation} \label{eq:tau_line}
\tau_{X,lu}(\nu) = \int_{0}^{\ell} n_{X,l} \ \sigma_{X,lu}(\nu,T)\  d\ell
.\end{equation}
In this expression, $n_{X,l}$ is the number density of scatterer $X$
in electronic state $l$.   For example, because the spontaneous
de-excitation time of \hi{} is very short, most \hi{} in astrophysical
media is in the ground state, and $n_{\ion{H}{i},1} = n_{\ion{H}{i}}$
is typically a very good approximation. The second term in
Eq.~\ref{eq:tau_line}, $\sigma_{X,lu}(\nu,T)$, is the average cross
section at frequency $\nu$ of a population of scatterers $X$ at
temperature $T$. This cross section is the convolution of the natural
Lorentzian line shape (in each scatterer's frame) with the Maxwellian
velocity distribution of scatterers, due to thermal and/or turbulent
motions. The mean thermal velocity of scatterers $X$ is
$\vth = (2k_BT/m_X)^{1/2}$, where $k_B$ is Boltzmann's constant and
$m_X$ the mass of $X$. We define the Doppler width
$\Delta\nu_D = (\vth/c)\nu_{lu}$, where $\nu_{lu}$ is the frequency of
the transition and $c$ the speed of light. The natural line width in
units of the Doppler width is $a = A_{ul} / (4\pi\Delta\nu_D)$, where
$A_{ul}$ is Einstein's coefficient for spontaneous emission. We can
then write the cross section with the usual Hjerting-Voigt function,
valid to first order in $(\mathrm{v}/c)$:
\begin{equation} \label{eq:sigma}
\sigma_{X,lu}(\nu,T) = \frac{\sqrt{\pi} e^2
  f_{lu}}{m_ec\Delta\nu_D} H(a,x).
\end{equation}
Here $e$ and $m_e$ are the electron's charge and mass, respectively;
$f_{lu}$ is the oscillator strength of the transition; the
dimensionless photon frequency 
$x = (\nu - \nu_{lu})/\Delta\nu_D$; and
\begin{equation}
H(a,x) = \frac{a}{\pi} \int_{-\infty}^{+\infty} \frac{e^{-y^2}}{(y-x)^2 + a^2}dy.
\end{equation}

This formalism can be used to compute the optical depth due to any
line. Only the values for $\nu_{lu}$, $A_{ul}$, and $f_{lu}$ need to
be changed for each transition. In Table~\ref{table:1} we provide
atomic data implemented in \ras{} for a selected sample of
lines.

\subsubsection{Dust optical depth}
\label{sec:tau_dust}

Along a path of length $\ell$ through a dusty medium, we can write
the total dust optical depth (due to both absorption and scattering)
as
\begin{equation}
\tau_{\rm dust}(\nu) = \int_0^\ell n_{\rm dust} \ \sigma_{\rm dust}(\nu) \ d\ell, 
\end{equation}
where $n_{\rm dust}$ and $\sigma_{\rm dust}$ are the number density
and cross section of dust grains. In \ras{} we follow by default the
formulation of \cite{laursen_ly_2009}: we define $\sigma_{\rm dust}$
as a cross section per hydrogen atom, and $n_{\rm dust}$ as a 
  pseudo number density given by
$n_{\rm dust} = (n_{\rm \ion{H}{i}} + f_{\rm ion} n_{\rm \ion{H}{ii}})
Z/Z_0$. Here, $f_{\rm ion}\sim 0.01$ is a free parameter describing
how much dust is present in ionised gas, and $Z_0 \sim 0.005 \ (0.01)$
is the mean metallicity of the Small and Large Magellanic Cloud (SMC and
LMC, respectively). As did  \cite{laursen_ly_2009}, we use the fits of
\citet{gnedin_escape_2008} to compute $\sigma_{\rm dust}(\nu)$ for
either the SMC or LMC models. Thanks to the modular style of \ras{}, it is
straightforward to implement alternative formulations for dust
opacity should it be required.

\subsection{Interactions with matter}
\label{sec:interactions}

When an interaction occurs, it may be with a dust grain, with
probability $P_{\rm dust} = \tau_{\rm dust} / \tau_{\rm tot}$, or the
photon packet may be absorbed in line $(X,lu)$, with probability
$P_{X,lu} = \tau_{X,lu}/\tau_{\rm tot}$.  The current version of
\ras{} implements three forms of interactions between photons and
matter: (1) resonant scattering (e.g. of \lya{} photons on \hi{}
atoms), (2) interactions with dust (either absorption or scattering),
and (3) transitions with multiple decay channels (e.g. fluorescent
lines of \ion{Fe}{ii}). We discuss how these are implemented in the
following  subsections.

\subsubsection{Resonant scattering}
\label{subsec:scatt_event}

In the frame of the gas, where particle motions are isotropic, a
scattering event will change the incoming frequency $\nu_{\rm in}$ and
direction of propagation $\vec{\mathrm k}_{\rm in}$ of the photon
packet into $\nu_{\rm out}$ and $\vec{\mathrm k}_{\rm out}$, depending
on the scatterer's velocity $\vec{\mathrm v}_X$ and mass $m_X$, so
that to first order in ${\mathrm v}/c$
\begin{equation} \label{eq:nuout_nuin}
  \nu_{\rm out} = \nu_{\rm in} \frac{1 + (\vec{\mathrm k}_{\rm
      out} - \vec{\mathrm k}_{\rm in}) \cdot \vec{\mathrm v}_X / c }
  { 1 + (1-\mu) \ (h\nu_{\rm in} /
      m_Xc^2)},
\end{equation}
where $\mu = \vec{\mathrm k}_{\rm out} \cdot \vec{\mathrm k}_{\rm in}$.
In Eq.~\ref{eq:nuout_nuin}, the denominator describes the recoil
effect \citep[e.g.][]{tasitsiomi_ly_2006}, which is generally small
\citep{adams_effect_1971}, while the numerator carries changes of
frames assuming coherent scattering in the frame of the
scatterer. Without loss of generality, we can choose a coordinate
system in which 
$\vec{\mathrm k}_{\rm in}=(1,0,0)$,
$\vec{\mathrm k}_{\rm out} = (\mu,\sqrt{1-\mu^2},0)$, and
$\vec{\mathrm v_X}/{\mathrm v}_{\rm th} = (u_\parallel, u_{\perp,1},
u_{\perp,2} )$, where $u_\parallel$ is the normalised velocity
component parallel to the incoming direction of
propagation. In this coordinate system, the outgoing frequency $\nu_{\rm out}$ is a function of
$u_\parallel$ and $u_{\perp,1}$ alone. For each scattering we draw a
value of $u_\parallel$ from a Gaussian distribution biased by the
prior that the scatterer interacts with a photon packet of frequency
$x_{\rm in} = (\nu_{\rm in}-\nu_{ul})/\Delta\nu_D$:
\begin{equation} \label{eq:uparallel_distribution}
f(u_\parallel) = \frac{a}{\pi H(a, x_{\rm in})} \frac{e^{-u_\parallel^2}}{a^2+(x_{\rm in}-u_\parallel)^2}.
\end{equation}
We then draw a value of $u_{\perp,1}$ from the unbiased Gaussian
velocity distribution of scatterers
$f(u_\perp) = e^{-u_\perp^2}/\sqrt{\pi}$.

The incoming and outgoing directions are related through a phase
function, and \ras{} implements a number of these functions. For \lya{}, for
example, we generally use two limiting phase functions, depending on
the frequency of the photons in the scatterer's frame
$\nu_{\rm scat,in} = \nu_{\rm in}(1- \vec{\mathrm k}_{\rm
  in} \cdot \vec{\mathrm v}_X / c)$, which reproduce the behaviour of
scatterings in the core of the line or in its
wings. Following
\citet{hamilton_directional_1940} and
\citet{dijkstra_polarization_2008}, we use the following phase
functions:
\begin{equation}\label{eq:Pcore}
P_{\rm core}(\mu) = 11/24 + 3 \mu^{2}/24
\end{equation}
when $|\nu_{\rm scat,in}-\nu_{X,lu}| < 0.2 \Delta\nu_D$, and 
\begin{equation}\label{eq:Pwing}
P_{\rm wing}(\mu) = 3(1+\mu^2)/8
\end{equation}
(i.e. Rayleigh scattering) otherwise. The transition
  at $0.2 \Delta\nu_D$ in the scatterer's frame is discussed in
  Appendix A of \citet{dijkstra_polarization_2008} and separates
  resonant and wing scattering
  events\footnote{Although related, this transition is
    not to be confused with the transition at $\sim3.3\Delta\nu_D$
    in the frame of the gas which separates the Gaussian core and
    Lorentzian wing of the absorption profile.}. For 
transitions other than \lya{}, we generally assume isotropic phase functions.

In environments with very high \ion{H}{i} opacities, \lya{} photons
will scatter many times locally until their frequency shifts enough
for them to make a larger spatial step \citep[see
e.g.][]{dijkstra_lyman_2014}. In order to reduce the computing time
used unnecessarily by these many scattering events, core-skipping
algorithms can be implemented, with which we can bias the frequency
redistribution in order to move the photon to the wing of the line
directly \citep{ahn_ly_2002,laursen_ly_2009}. We   implemented in
\ras{} the core-skipping algorithm described in
\citet[][Sect. 3.2.4]{smith_lyman_2015}.  We note that this
acceleration scheme is accurate for media without dust, but will lead
to a small underestimation of extinction when dust is present (see
discussion in \citealt{laursen_ly_2009}). We do not use this
acceleration in this paper, except in Sect.~\ref{sec:lyaExample}.

\subsubsection{Interactions with dust}
\label{subsec:scatt_event_dust}

When a photon packet interacts with a dust grain, it may either be
absorbed or scattered with a probability set by the dust albedo
$a_{dust}(\nu)$. In \ras{} the dust albedo is a free parameter. By
default we use $a_{dust}=0.32$ (i.e. $32\%$ of the photons are
reflected) for \lya{} (see \citet{li_infrared_2001} for other
frequencies).

When a photon packet scatters on a dust grain, we use the
Henyey-Greenstein \citep{henyey_diffuse_1941,laursen_ly_2009} phase
function to compute its outgoing direction
\begin{equation}\label{eq:Pdust}
P_{HG}(\mu) = \frac{1}{2} \frac{1-g^2}{(1+g^2-2g\mu)^{3/2}},
\end{equation}
with $g$ the asymmetry parameter. As for the dust albedo,
this parameter is a function of frequency, and a free parameter in
\ras{} (with a default value $g=0.73$ for \lya{}, taken from
\citealt{li_infrared_2001}).

\subsubsection{Fluorescent lines}

A number of transitions of great observational interest originate from
ions that have more complex level structures than simple resonant
transitions, in the sense that one absorption channel, say from level
$l_1$ to level $u$, leads to more than one decay channel, down to
levels $l_1, l_2, ...$. We refer to these transitions of type $ul_2$
as fluorescent, and a few examples are given in
Table~\ref{table:1}. \ras{} deals with these lines easily by having one
module per absorption channel. The absorption of photon packets
is computed exactly as for resonant lines, but re-emission has a
probability of being non-resonant according to the following ratio of
Einstein coefficients:
\begin{equation}
P_{ul_i} = \frac{A_{ul_i}}{\sum_{i} A_{ul_i}}.
\end{equation}
In the case of fluorescent re-emission, \ras{} simply casts a photon with
$\nu_{ul_i}$ in the frame of the scatterer, assuming an isotropic
phase function. When the decay channel is resonant, we follow
Sect.~\ref{subsec:scatt_event}.

\subsection{Numerical implementation}
\label{sec:mcrt}

\subsubsection{Phase functions}
\label{subsec:phase_functions}

Following \citet{barnes_PhD_2009}, the phase functions given in
Eqs.~\ref{eq:Pcore} and \ref{eq:Pwing} are analytically integrable and
invertible. The solution of this cubic polynomial is a function of
the form
\begin{equation}
  \mu = (A+B)^{1/3} - (A-B)^{1/3}
\end{equation}
with
\begin{eqnarray}
  B_{\rm core} = 6(2r-1) &\textrm{and}& A_{\rm core} = \sqrt{B_{\rm core}^2+\frac{11^3}{27}}~,\\
  B_{\rm wing} = (4r-2)  &\textrm{and}& A_{\rm wing} = \sqrt{B_{\rm wing}^2+1}~,
\end{eqnarray}
and $r$ a univariate between 0 and 1.  The Henyey-Greenstein phase
function (given in Eq.~\ref{eq:Pdust}) is also analytically integrable
and invertible, so we obtain $\mu$ values sampling this distribution
by drawing univariate $r$ in $[0;1]$ and computing
\begin{equation}
\mu = \frac{1}{2g} \left( 1+g^2 - \frac{1-g^2}{1-g+2gr} \right).
\end{equation}

\subsubsection{Voigt function} \label{sec:voigt}

\begin{figure}
  \centering
  \includegraphics[width=9cm]{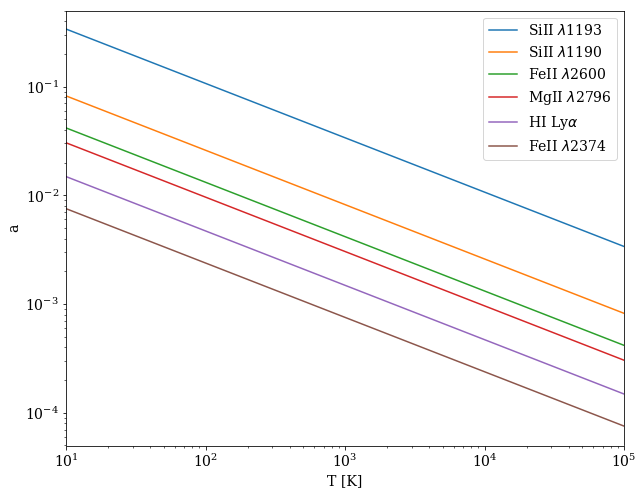}
  \caption{Voigt parameter ($a = A_{ul} / (4\pi\Delta\nu_D)$) as a
    function of gas temperature for various species and/or transitions. Some species may not exist in the full temperature range. }
  \label{fig:aT}
\end{figure}

In Sect.~\ref{sec:tau_line} (Eq.~\ref{eq:sigma}), we see that we
needs to evaluate the Hjerting-Voigt function $H(a,x)$ in order to
compute the line optical depth. There is no analytic solution for this
integral, and an accurate numerical integration is computationally
expensive. This operation is one of the most frequent  in \ras{},
and thus it is essential to use efficient approximations. Unlike most
codes, which only  focus on the \lya{} line
\citep[e.g.][]{tasitsiomi_ly_2006,smith_lyman_2015}, \ras{} deals with
other lines, which have different values, at a given temperature, of
the Voigt parameter $a$. The normalisation of $a$, at a given
temperature, is proportional to $A_{ul} m_X^{1/2}$. Figure~\ref{fig:aT}
shows $a-T$ relations for different  lines selected from
Table~\ref{table:1}. For some transitions, the values of $a$ can
easily be more than one order of magnitude higher than for \lya{}. 

In \textsc{rascas}, we implement three different approximations taken
from the literature. The simplest approximation we implement is that
introduced by \citet{tasitsiomi_ly_2006} in their Eqs. 7 and 8.  The
second approximation we implement is the more elaborate method by
\citet{smith_lyman_2015}, given in their Appendix A1 (their
Eq. A1). The third option is the implementation of the rational form
given by \citet{humlicek_optimized_1982}, which they provide as a
Fortran routine (called \texttt{W4}) in the appendix of their paper.

\begin{figure*}
  \centering
  \includegraphics[width=18cm]{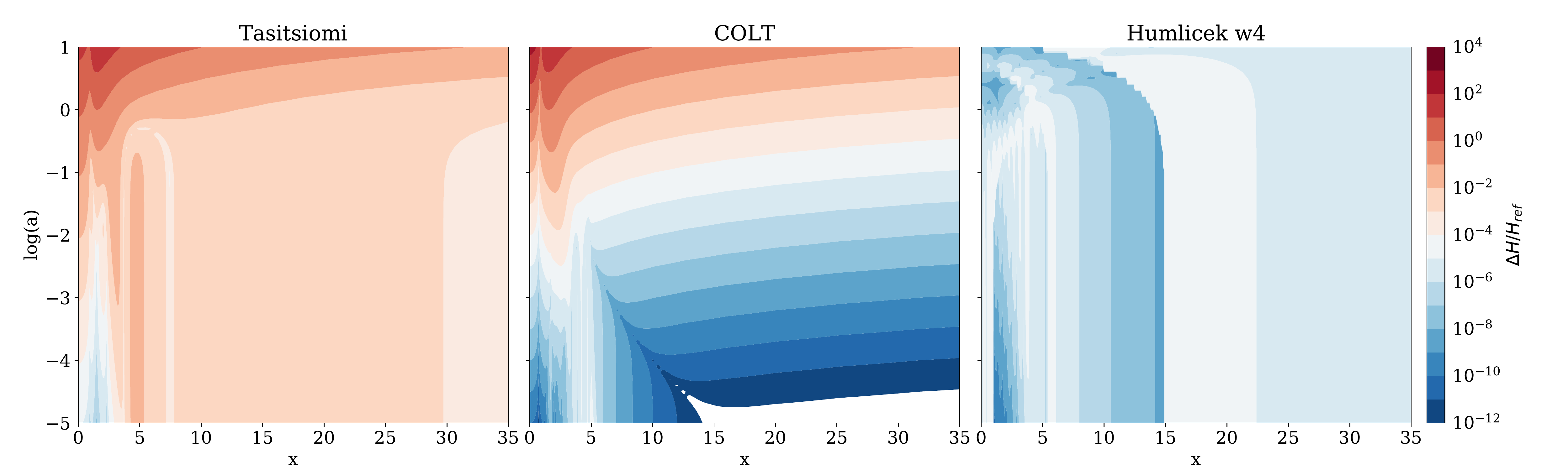}
  \caption{Contour plot of the relative errors of the three
    approximations for the Voigt function implemented in
    \ras. \textit{Left:} Approximation from
    \citet{tasitsiomi_ly_2006}. \textit{Middle:} Approximation from
    \citet{smith_lyman_2015}. \textit{Right:}  \texttt{W4}
    approximation from \citet{humlicek_optimized_1982}. Bluer means
    more accurate, while redder means less accurate. White parts have
    a relative error better than $10^{-12}$. The transition between
    light blue and light red is at $10^{-4}$. }
  \label{fig:voigt_accuracy}
\end{figure*}

In Fig.~\ref{fig:voigt_accuracy} we compare the accuracy of these three
approximations over a wide range of $a$ and $x$ values.  We evaluate
each method by comparing their predictions to an accurate reference
given by the scipy function \texttt{wofz}\footnote{Steven G. Johnson,
  Faddeeva W function
  implementation. \texttt{http://ab-initio.mit.edu/Faddeeva}}, which computes
the Faddeeva function for complex argument, whose real part is the $H$
function. It has been shown that this implementation has an accuracy
to least 13 significant digits in both the real and imaginary
parts \citep[e.g.][]{oeftiger_review_2016}. As already shown by
\citet{schreier_optimized_2011} and
\citet{schreier_computational_2017}, we find that the Humlicek
\texttt{W4} form is very accurate all over the $(a,x)$ domain, with a
relative error lower than $10^{-4}$, as shown by the authors. The
approximation given by \citet{tasitsiomi_ly_2006} and widely used for
\lya\ RT in astrophysics \citep[e.g.][]{verhamme_3d_2006,
  semelin_lyman-alpha_2007, abe_seurat:_2018} has an accuracy of around
$1\%$ for the \lya\ line at $T=10^4$~K. This accuracy degrades for
higher values of $a$. In comparison, the approximation proposed by
\citet{smith_lyman_2015} has an accuracy lower than $10^{-4}$ for
$a < 10^{-2.5}$, i.e. for \lya\ at $T > 500$~K. However this
approximation gives less accurate results at $a \ga 10^{-2}$.

\begin{figure}
  \centering
  \includegraphics[width=9cm]{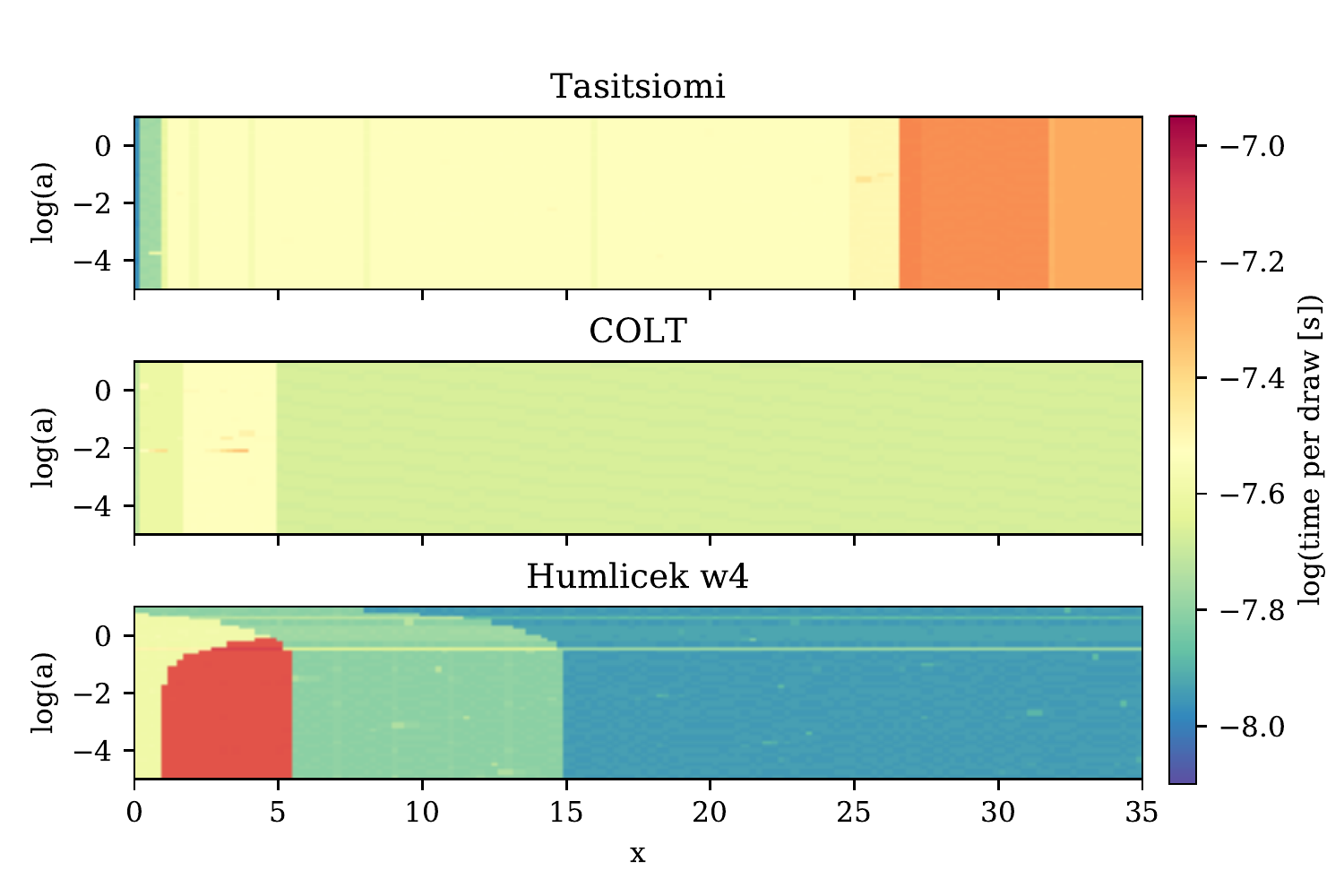}
  \caption{Measured performance for the computation of the $H(a,x)$
    function in the $x - \log a$ plane for the three approximations
    implemented in \ras. Colour-coding indicates the log of the time per call
    in seconds, where bluer indicates faster times.}
  \label{fig:voigt_perf}
\end{figure}

In Fig.~\ref{fig:voigt_perf}, we also compare the performance of the
three approximations. We time each function by calling it one million
times for each point in a grid of values in the $(\log_{10}(a),x)$
plane, and by measuring the mean CPU time per draw. The top panel
shows the timing of the approximation of
\citet{tasitsiomi_ly_2006}. The fast region below $x\sim0.92$ is where
the approximation relies only on an exponential. At higher $x$ values
the approximation also requires the evaluation of an extra term, and is
thus a bit slower. It is not clear why the method becomes slower at
very high values of $x$, and this may be tied to uncontrolled
compiler behaviours. In the middle panel, the three regions of the
approximation of \citet{smith_lyman_2015} are clearly visible, even
though the CPU time is  homogeneous across the full range of
values explored here. In the bottom panel, we see clearly the four
regions of the method of \citet{humlicek_optimized_1982} implemented
in the \texttt{W4} function. The slowest region, at low $x$ and
$\log a$, is due to a combination of an exponential function in the
complex plane and a fraction of two polynomial expressions. At $x > 5$
their method is extremely fast,  and is faster than the other methods we
tested.

Another way to benchmark the three approximations in order to avoid
cache effects and some compiler optimisations is to draw a random
distribution of $x$ values, and to compute $H(x,a)$ for a fixed value
of $a$. Here we take $a = 4.72\times 10^{-3}$, which corresponds to
$T=10^4$~K for the \lya\ line and we draw $10^8$ values of $x$
randomly distributed in the range $0 < x < 35$. We find that this
takes $3.924 \mathrm{s}$ with the \citet{tasitsiomi_ly_2006}
approximation, $2.188 \mathrm{s}$ with the \citet{smith_lyman_2015}
approximation, and $2.612 \mathrm{s}$ with the
\citet{humlicek_optimized_1982} approximation.

In conclusion, the three approximations have different behaviours
both in terms of accuracy and in terms of computational cost. Choosing
one method over another depends on the compromise between performance
and accuracy that a problem set requires. The
\citet{humlicek_optimized_1982} approximation is clearly better than
the others in terms of accuracy since it has a $10^{-4}$ accuracy  
over a huge range of $x$ and $a$. However, in terms of performance,
it depends strongly on the domain of $x$ and $a$, and may be a few
times slower than the others at most. On the other hand, the
approximation proposed by \citet{smith_lyman_2015} appears much more
homogenous and seems a bit faster for a long series of computations
sampling $x$ values uniformly. In a typical experiment, however, most
draws will happen near $x\sim0$, and the method of
\citet{tasitsiomi_ly_2006} may turn out to be faster, though less
accurate. We would generally recommend using the method of
\citet{humlicek_optimized_1982} for metal lines, which may lead to
high values of $a$. This guarantees a good level of precision and
the slight overhead is certainly affordable as metal-line photons do
not generally undergo many scatterings. In the case of \lya\ line
transfer, where MCRT is expensive due to the huge number of
scatterings that happen preferentially at low $x$, the
\citet{smith_lyman_2015} approximation appears to be a good
compromise.

\subsubsection{Generating $\upar$}
\label{sec:upar}

\begin{figure}
  \centering
  \includegraphics[width=9cm]{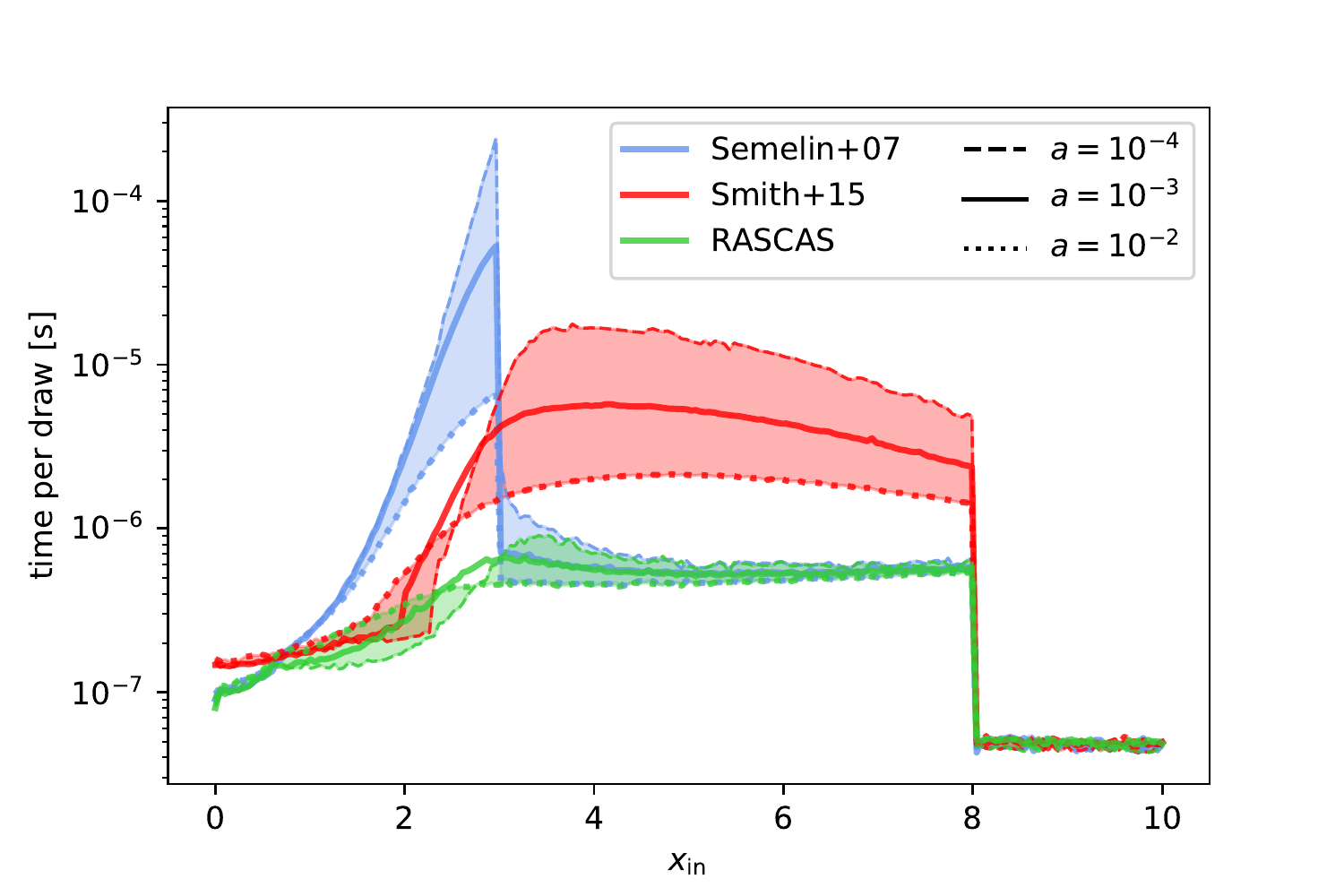}
  \caption{Measured performance of our draws of $\upar$ in
      terms of CPU time per draw vs. input frequency $x_{\rm
        in}$. The methods of \citet{semelin_lyman-alpha_2007} and
      \citet{smith_lyman_2015} are shown in blue and red, while our
      approach is shown in
      green. The solid lines are obtained with $a=10^{-3}$, while the
      dotted and dashed lines correspond to $a=10^{-2}$ and
      $a=10^{-4}$. The shaded areas indicate the area between these two
        curves. The non-monotonic behaviour of CPU cost with
      $a$ for our method and that of \citet{smith_lyman_2015} can be seen: at low
      $x_{\rm in}$, low $a$ values are the cheapest, while at high
      $x_{\rm in}$, low $a$ values are the most expensive draws.}
  \label{fig:upar_perf}
\end{figure}

We implement the rejection method of \citet{zheng_monte_2002} to
sample the distribution of $u_\parallel$ given in
Eq.~\ref{eq:uparallel_distribution}. We use the piecewise comparison
function:
\begin{equation}
 g(u_\|) \propto
   \begin{cases}
     g_1 = 1 / \left[ a^2 + (x-u_\|)^2 \right] & \; u_\| \leq u_0 \\
     g_2 = e^{-u_0^2} / \left[ a^2 + (x-u_\|)^2 \right] & \; u_\| > u_0
   \end{cases}
,\end{equation}
where $u_0$ is a separation parameter and the corresponding acceptance
fraction is $\exp(-u_\|^2)$ for $g_1$ and $\exp(u_0^2-u_\|^2)$ for
$g_2$. The efficiency of this method (i.e. the number of rejections)
depends heavily on the value of $u_0$. \citet{semelin_lyman-alpha_2007} use
an empirical fit (their Eq. 17) at $x > 3$ and set $u_0=0$ at
$x\leq3$. \citet{smith_lyman_2015} propose an elegant method used to derive
analytic estimates of $u_0$ for core or wing scatterings (their
Eqs. 31 and 32), and effectively set $u_0=0$ at $x\leq 1$. Here, we use
a 2D polynomial function $u_0(x,a),$ which we obtained by empirically
finding the $u_0$ values producing the fastest execution time for a
grid of values $(x,a)$. This fitting function is the following:
\begin{eqnarray} \nonumber
u_0 &=& 2.648963+2.014446 \zeta+0.351479\zeta^2 \\ \nonumber 
 & &+ x (-4.058673-3.675859\zeta-0.640003\zeta^2 \\ \nonumber 
 & &+ x (3.017395+2.117133\zeta+0.370294\zeta^2 \\ \nonumber 
 & &+ x (-0.869789-0.565886\zeta-0.096312\zeta^2 \\ \nonumber 
 & &+ x (0.110987+0.070103\zeta+0.011557\zeta^2 \\ 
 & &+ x (-0.005200-0.003240\zeta-0.000519\zeta^2))))),
\end{eqnarray} 
where $\zeta = \log_{10}(a)$. At high values of $x$ ($x\geq 8$ by
default), we follow \citet{smith_lyman_2015} and directly sample a
Gaussian distribution centred at $1/x$.

In Fig. \ref{fig:upar_perf} we compare the performance  of our
implementation with those of \citet{semelin_lyman-alpha_2007} and
\citet{smith_lyman_2015}\footnote{To be fair, we compare our
  implementation of \citet{smith_lyman_2015}, not theirs.}. For $x<1$,
all methods are equally good, and the results there do not depend much
on the value of $u_0$. At $1<x<3$, the method of
\citet{smith_lyman_2015} improves over that of
\citet{semelin_lyman-alpha_2007} by about an order of magnitude, but
the time per draw still drifts up by more than one order of magnitude
towards $x=3$. At $x>3$, the method of
\citet{semelin_lyman-alpha_2007} is one or two orders of magnitude
faster than that of \citet{smith_lyman_2015} depending on the value of
$a$ (low values of  $a$ are slower). At $x>8$, the methods change as we draw
$\upar{}$ from a Gaussian distribution directly. The polynomial
function that we implement to decide the value of $u_0$ as a function
of $(x,a)$ produces results that are always better than or equal to the results of both the
\citet{semelin_lyman-alpha_2007} and \citet{smith_lyman_2015} methods.

\subsection{Test cases}
\label{test_cases}

In this section, we present a series of tests that were carried out to
validate the MC implementation of the radiative transfer in
\textsc{rascas}. These tests consist of numerical experiments of 
single-scattering events and of the full propagation of photons in
idealised geometries, for which known analytic solutions exist.

\subsubsection{Single-scattering experiments}

\begin{figure}
  \centering
        \hspace{-0.8cm}
  \includegraphics[width=8.4cm]{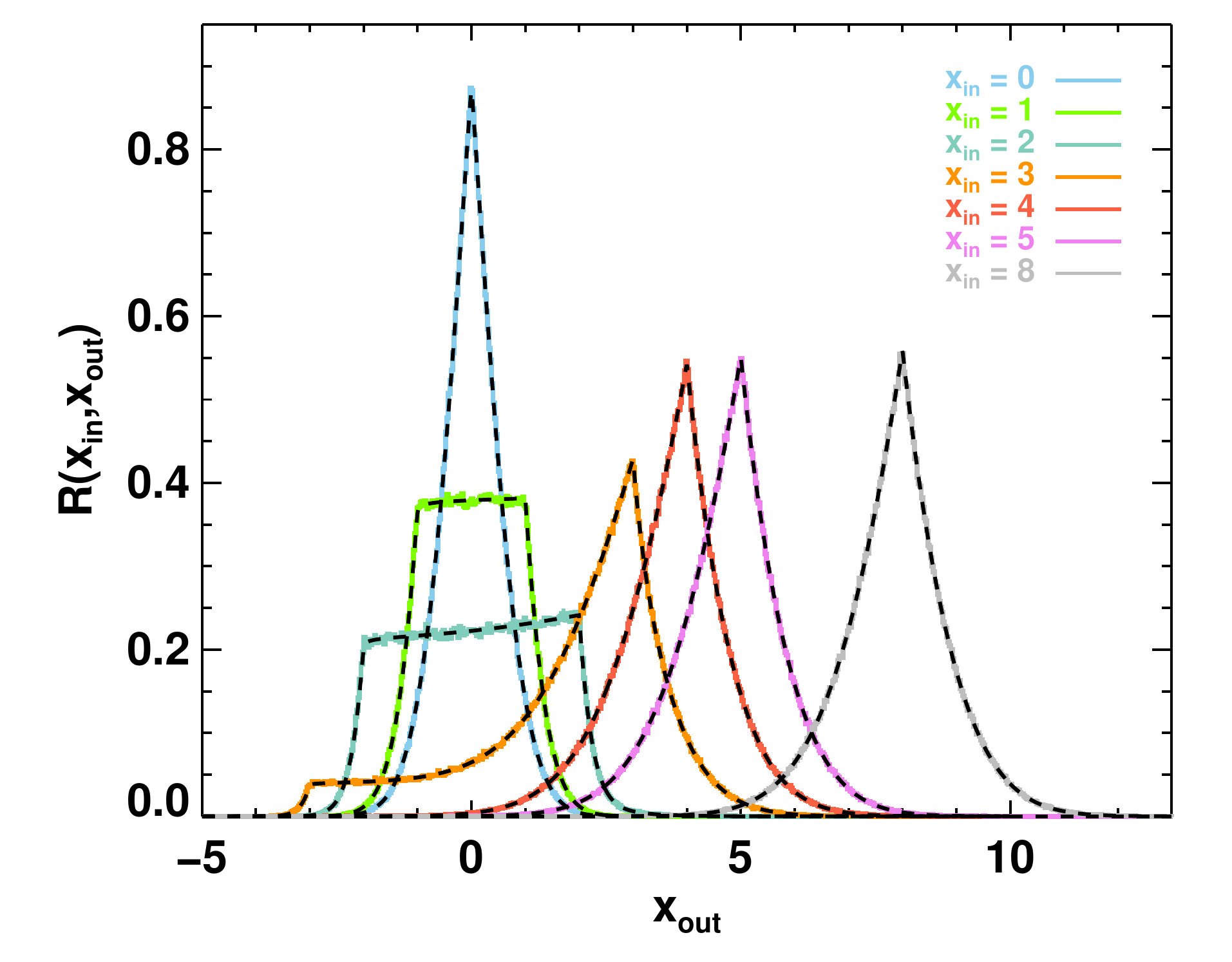}
  \includegraphics[width=8.1cm]{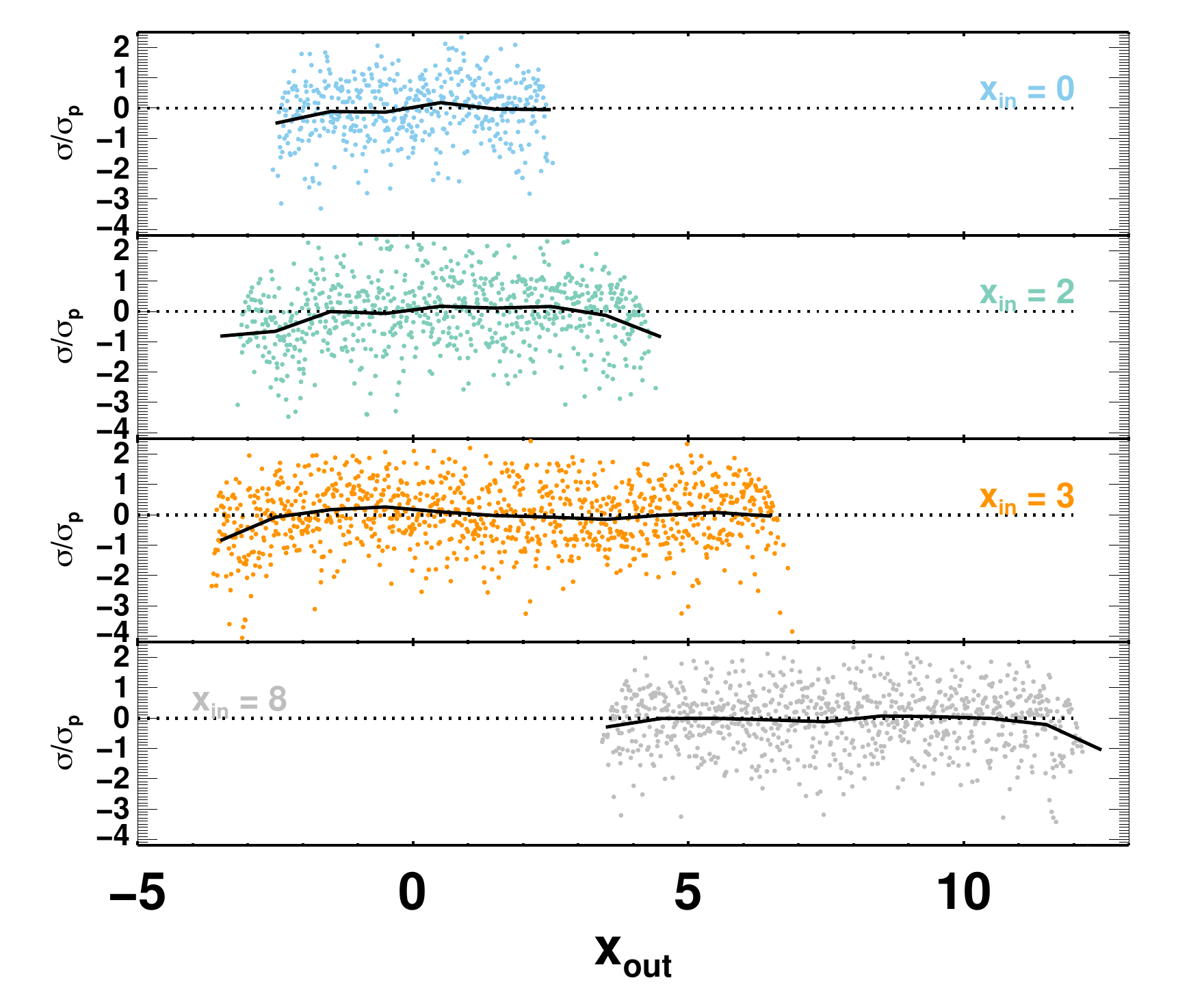}
  
  \caption{\textsc{rascas} simulations of single-scattering events in
    a \ion{H}{I} medium with $T=100$~K assuming isotropic angular
    redistribution and no recoil effect. \textit{Top}: Frequency redistribution, $R(x_{\rm in},x_{\rm out})$,
    in Doppler units ($x_{\rm out}$) of Lyman$-\alpha$ photons emitted
    at $x_{\rm in}$ after one single-scattering on a hydrogen
    atom. The solid coloured curves are the results from
    \textsc{rascas} assuming $x_{\rm in}=0, 1, 2, 3, 4, 5$, and 8, while
    the dashed black lines are the analytic solutions of
    \citet{hummer_non-coherent_1962}. \textit{Bottom}: Accuracy of
    $R(x_{\rm in},x_{\rm out})$ from \textsc{rascas} compared to the
    solution of \citet{hummer_non-coherent_1962} as a function of
    $x_{\rm out}$. Here $\sigma$  is the relative error between
    $R(x_{\rm in},x_{\rm out})$ obtained from \textsc{rascas} and the
    solution of \citet{hummer_non-coherent_1962}, and $\sigma_{\rm p}$ is
    the relative Poisson error in each bin of $x_{\rm out}$ due the
    limited number of photons used in this simulation
    ($1/\sqrt{N_{\rm phot}}$). In each sub-panel, the solid black
    curves show the mean value of $\sigma / \sigma_{\rm p}$.}
         \label{indiv_scatt}
   \end{figure}

In Fig.~\ref{indiv_scatt} (top panel) we show the frequency
redistribution ($x_{\rm out}$) of Lyman$-\alpha$ photons emitted at
various input frequencies ($x_{\rm in}$) after one scattering on
hydrogen atoms. Here we assumed a temperature of 100~K and isotropic
angular redistribution ($W(\theta)=$const.), and we neglected the
recoil effect. Our results and the exact redistribution functions
derived by \citet{hummer_non-coherent_1962}  are
nearly indistinguishable. To assess the reliability of the agreement
between \textsc{rascas} and the \citet{hummer_non-coherent_1962}
solution, we plot the relative error between the two ($\sigma$) in
units of the relative Poisson error (i.e. the variance due to the
limited number of photons per bin in the simulation; $\sigma_{\rm p}$)
in the bottom panel of Fig.~\ref{indiv_scatt}.  We find that the
redistribution functions as computed by \textsc{rascas} agree almost
perfectly with the exact solution, and the tiny differences between the
two are due to statistical noise. The overall agreement of \textsc{rascas} with the
\citet{hummer_non-coherent_1962} solutions confirms the validity of our
implementation of the atomic physics for the different transitions
available, and of the $\upar$ rejection method used to determine
the scatterer's velocity along the propagation of the photons (see
Sect.~\ref{sec:upar}). We show a few additional tests
  in Appendix~B which further validate the implementation of atomic
  physics in \ras{}.

\subsubsection{Idealised configurations}

In this section we first compare simulations of Lyman$-\alpha$ RT in
idealised static plane-parallel slabs of \ion{H}{I} with the analytic
solutions of \citet{neufeld_transfer_1990} (which are based on the
work by \citealt{harrington_scattering_1973}). To derive their
solutions, these authors assume that photons scatter mostly in the
wings with an absorption Voigt profile approximated as a Lorentzian
profile ($\Phi(x) \approx a / \pi x^2$). Their formula are therefore
expected to be exact only at very large \ion{H}{I} optical depth and
low temperature (i.e. in the extremely optically thick regime
($a\tau_{\rm \hi} \gtrsim 10^3$)\footnote{In the following, unless
  specified otherwise, $\tau_{\rm \hi}$ stands for
  $\tau_{\rm \hi, Ly\alpha}$}).

\begin{figure*}
  \centering
  \hspace{-0.8cm}
  \includegraphics[width=5.75cm]{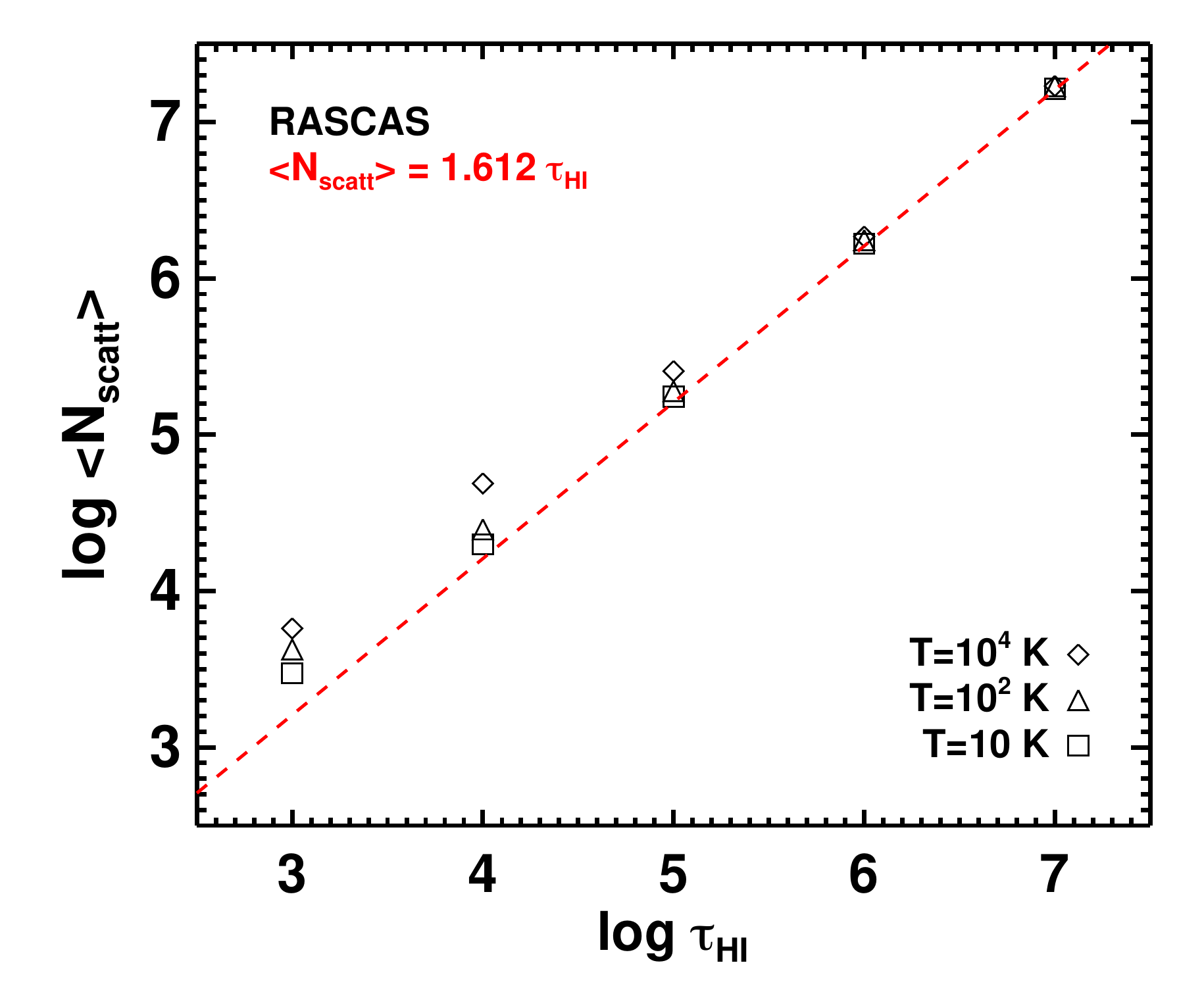}           
  \includegraphics[trim=0.75cm 5.9cm 0.5cm 7cm,clip=true, width=6.25cm]{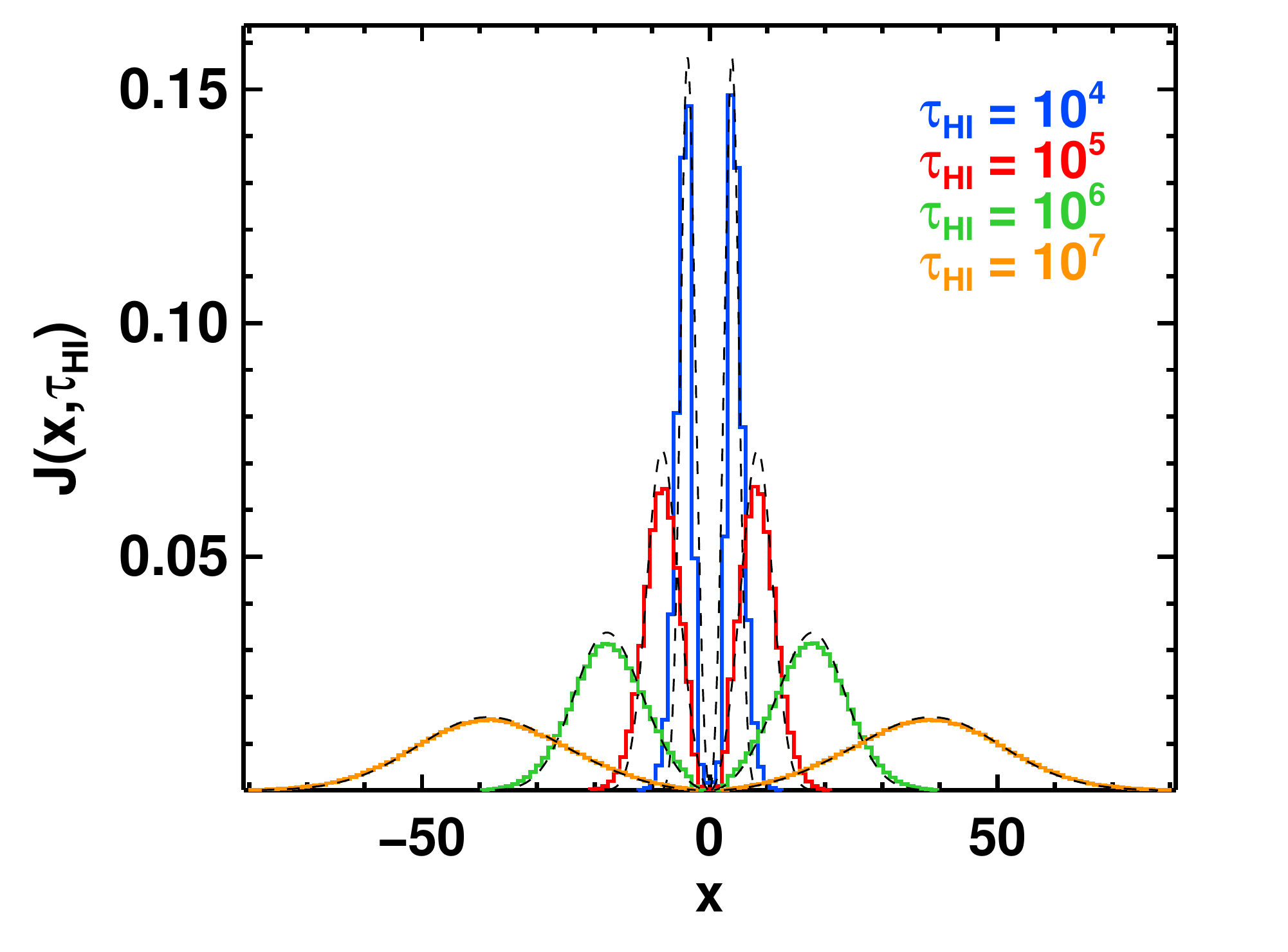}  
  \includegraphics[trim=1.5cm 5.9cm 0.5cm 7cm, clip=true, width=6.cm]{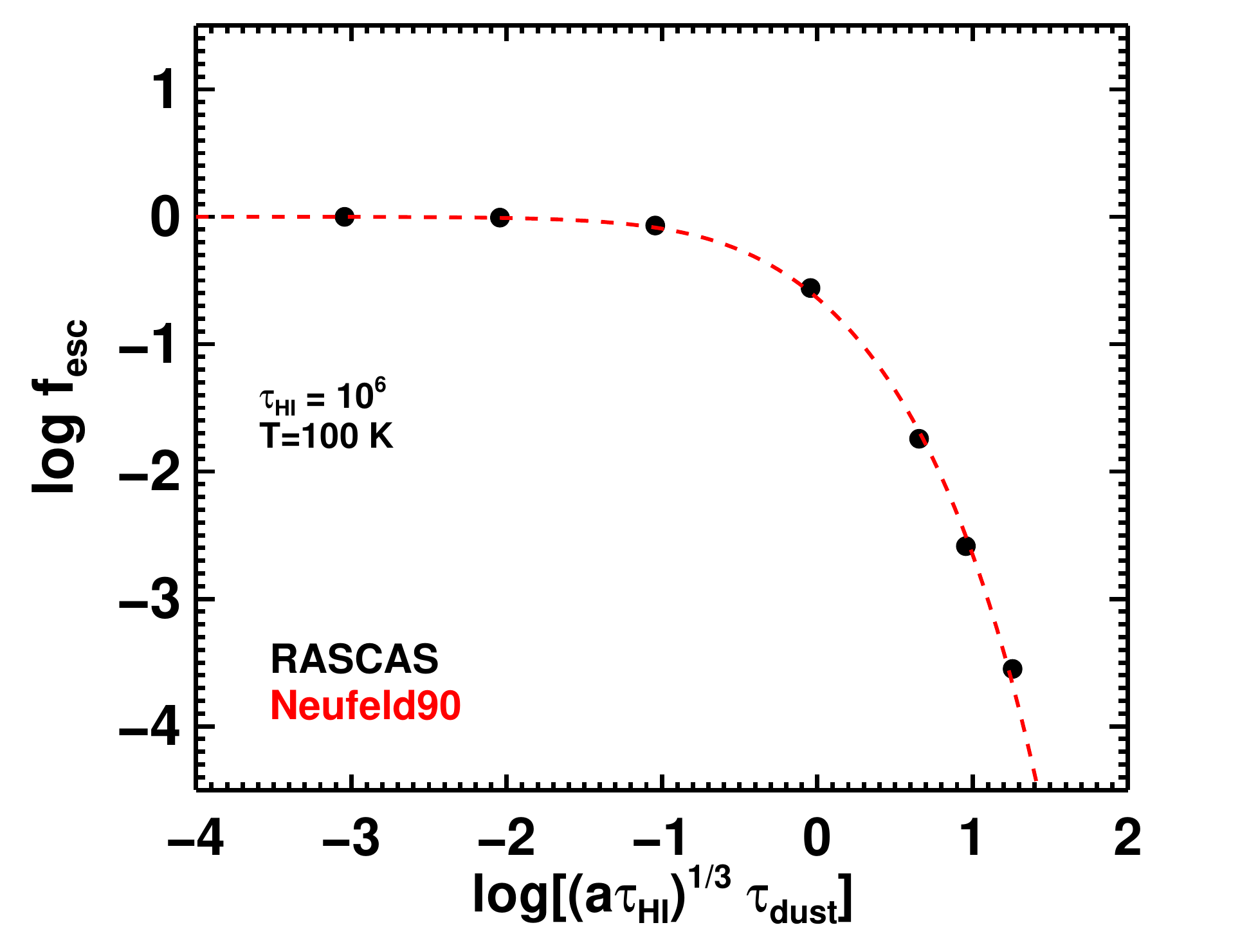}
\vspace{0.1cm}
      \caption{\textsc{rascas} simulations of the transfer of 
        Lyman$-\alpha$ photons ($x_{\rm in}=0$) emitted at the origin
        of a uniform plane-parallel slab. \textit{Left}: Mean number
        of scatterings until escaping the slab as a function of the
        vertical \ion{H}{I} opacity $\tau_{\rm \hi}$. \textsc{rascas}
        runs are shown by black symbols for different assumed gas
        temperatures ($T=10, 10^{2}, 10^{4}$~K). The red dashed line
        corresponds to the analytic approximation found by
        \citet{harrington_scattering_1973}, $N_{\rm scatt} = 1.612
        \tau_{\rm \hi}$. \textit{Middle}: Variation of the emergent
        spectrum as a function of $\tau_{\rm \hi}$. The solid coloured
        lines are the results from \textsc{rascas}, while the black
        dashed curves are the analytic predictions of
        \citet{neufeld_transfer_1990}. We assume $T=100$~K, no recoil effect,
        and an isotropic angular redistribution. \textit{Right}: Evolution of
        the Lyman$-\alpha$ escape fraction $f_{\rm esc}$ as a function
        of dust optical depth ($\tau_{\rm dust}$) in the case of a
        slab with $\tau_{\rm \hi} = 10^6$ and $T=10^2$~K. The results
        from \textsc{rascas} (filled circles) reproduce very well the
        analytic prediction of \citet{neufeld_transfer_1990} (red
        dashed line).} 
\label{slab}
\end{figure*}

In Fig.~\ref{slab} (left panel) we show the mean number of
scatterings of Lyman$-\alpha$ photons ($x_{\rm in}=0$) as a function
of $\tau_{\rm \hi}$ for three different slab temperatures
($T=10, 10^{2}, 10^{4}$~K, or equivalently
$a \approx 0.0149, 0.0047, 0.00047$). As expected, we see that our
simulations (black points) only reach an excellent agreement with the
analytic formula derived by \citet{harrington_scattering_1973} (red
dashed curve) at low $T$ and high \ion{H}{I} opacities. As shown by
\citet{neufeld_transfer_1990}, the emergent spectrum for the static
slab configuration is a double-peak profile centred on
$x_{\rm out}=0$. The middle panel of Fig.~\ref{slab} shows the
spectra computed with \textsc{rascas} for various \ion{H}{I} opacities
(coloured solid curves) assuming $x_{\rm in}=0$ and $T=100$~K. Again,
our numerical predictions are closer to the
\citet{neufeld_transfer_1990} solutions when the medium is optically
thicker ($\tau_{\rm \ion{H}{i}} \gtrsim 10^6$).

Finally, using the same framework, \citet{neufeld_transfer_1990}
investigated the radiative transfer of Lyman$-\alpha$ photons in an
absorbing medium and derived an approximated formula for the escape
fraction $f_{\rm esc}$ of photons through a dusty slab. Because of
resonant scattering, the dust attenuation of the Lyman$-\alpha$ line
is enhanced and not only depends on the dust opacity
$\tau_{\rm dust}$, but also on the \ion{H}{I} optical depth
$\tau_{\rm \ion{H}{i}}$ and the gas temperature (via the parameter
$a$). The right panel of Fig.~\ref{slab} shows $f_{\rm esc}$ as a
function of $(a\tau_{\rm \hi})^{1/3} \tau_{\rm dust}$ assuming
$\tau_{\rm \hi} = 10^6$ and $T=10^2$~K for several \textsc{rascas}
runs (black points) compared to the formula of
\citet{neufeld_transfer_1990} (red dashed line). We find that the
escape fraction decreases in a non-linear fashion as $\tau_{\rm dust}$
increases, in good agreement with the analytic prediction.

\begin{figure}
  \centering
  \includegraphics[width=8.2cm]{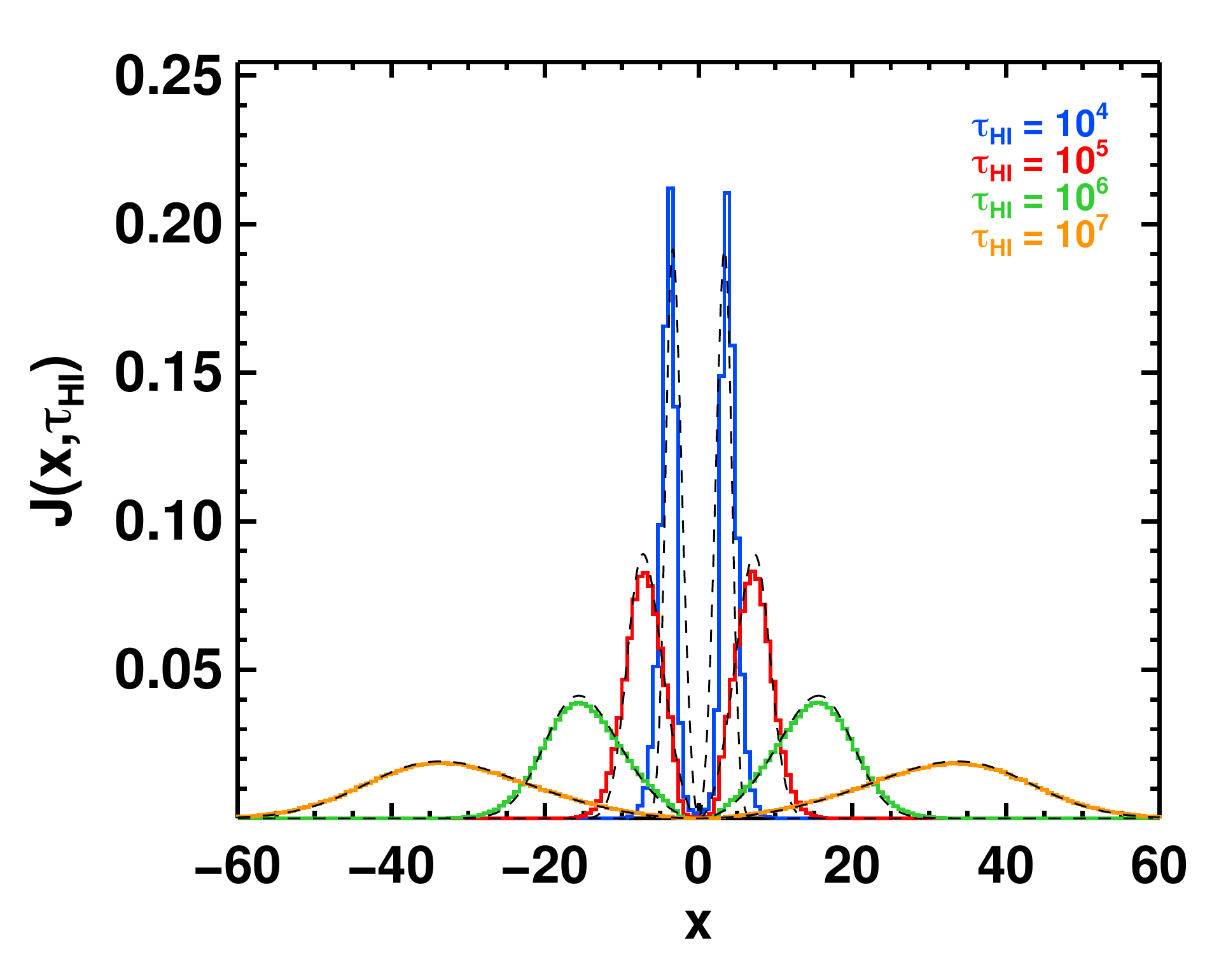}
  \includegraphics[width=8.2cm]{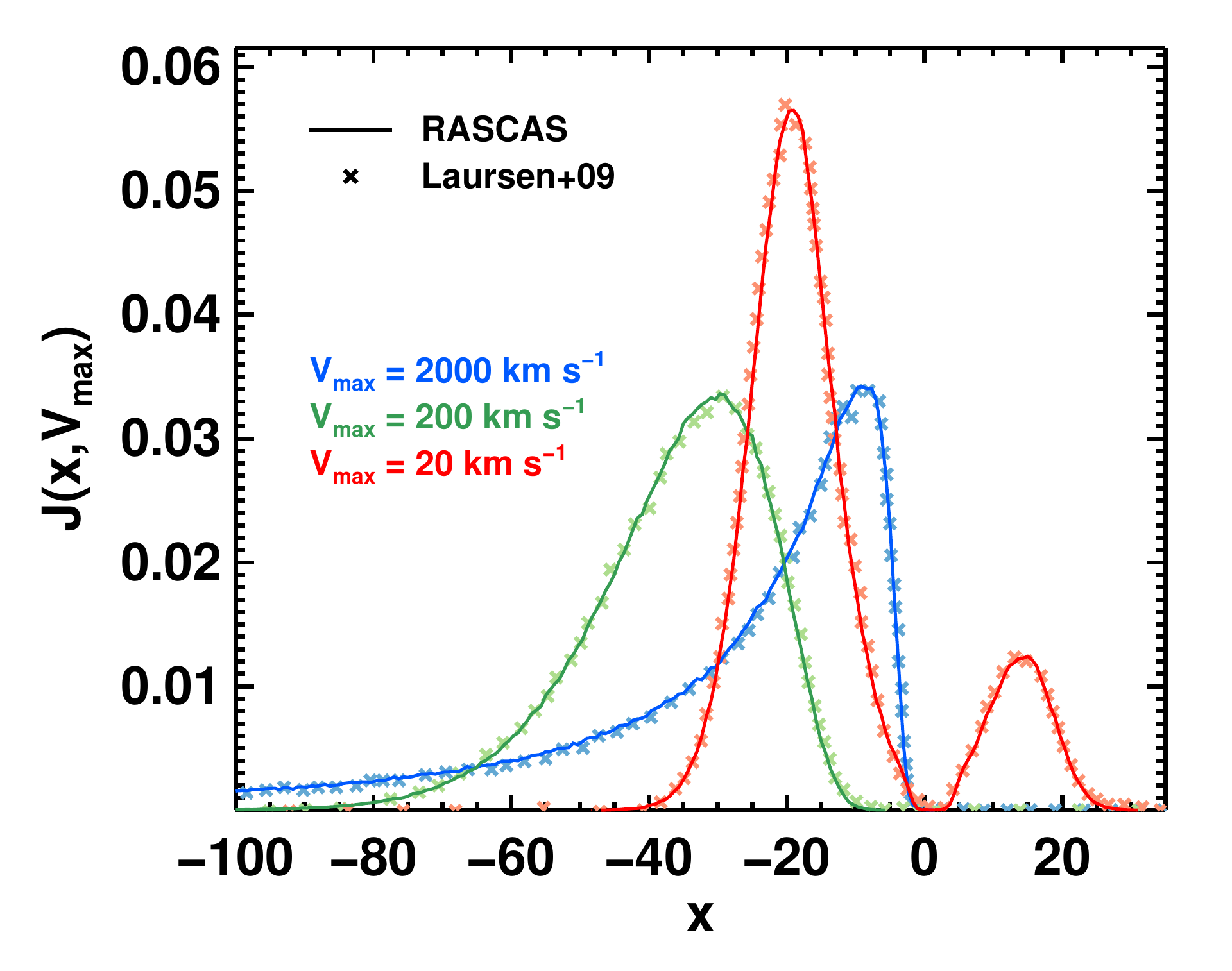}
  \caption{\textsc{rascas} simulations of the transfer of
    Lyman$-\alpha$ photons ($x_{\rm in}=0$) emitted at the centre of a
    uniform sphere at $T=10^2$~K. \textit{Top}: Variation of the
    emergent spectrum as a function of $\tau_{\rm \hi}$ for a static
    sphere. The solid coloured lines are the results from
    \textsc{rascas}, while the black dashed curves are the analytic
    solutions of \citet{dijkstra_ly_2006}. \textit{Bottom}: Non-static
    sphere with a \hi{} column density of $2\times 10^{21}$ cm$^{-2}$ in
    which the gas velocity increases linearly with radius from 0 at
    the centre to $V_{\rm max}$ at the edge of the outflow. The
    spectra predicted by \textsc{rascas} (curves) are compared to
    \citet{laursen_ly_2009}, who did the same experiments with their
    code (crosses).}
\label{sphere}
\end{figure}

Various authors have investigated the Lyman$-\alpha$ RT in other
simple configurations \citep{loeb_scattered_1999, dijkstra_ly_2006,
  laursen_ly_2009}. In Fig.~\ref{sphere} (top panel), we show the
spectra emerging from a uniform static sphere, and compare them with
the analytic solution derived by \citet{dijkstra_ly_2006} who used an approach
similar  to that of  \citet{neufeld_transfer_1990}. Again, we confirm
that \textsc{rascas} recovers  the expected line profile well,
especially in the very optically thick regime
($\tau_{\rm \hi} \gtrsim 10^6$). In order to test our code in a
non-static experiment, we also perform the simulations of
\citet{laursen_ly_2009} since no analytic solution exists for
Lyman$-\alpha$ propagation in moving media. The bottom panel of
Fig.~\ref{sphere} shows the spectra for three homogeneous spherical
outflow models at $T=10^2$~K where the gas velocity increases linearly
with radius from 0 to a maximum velocity $V_{\rm max}$. For high
outflow velocities ($V_{\rm max}=200, 2000$ km s$^{-1}$), the
resulting line profiles are asymmetric and are shifted towards negative x
(i.e. longer wavelengths), while for $V_{\rm max}=20$ km s$^{-1}$ a
fraction of the photons can escape with positive x values
(i.e. blueward of the line centre). In all cases we see that the
agreement between \textsc{rascas} (curves) and \citet{laursen_ly_2009}
(crosses) is excellent.

\begin{figure}
  \centering
  \includegraphics[width=9cm]{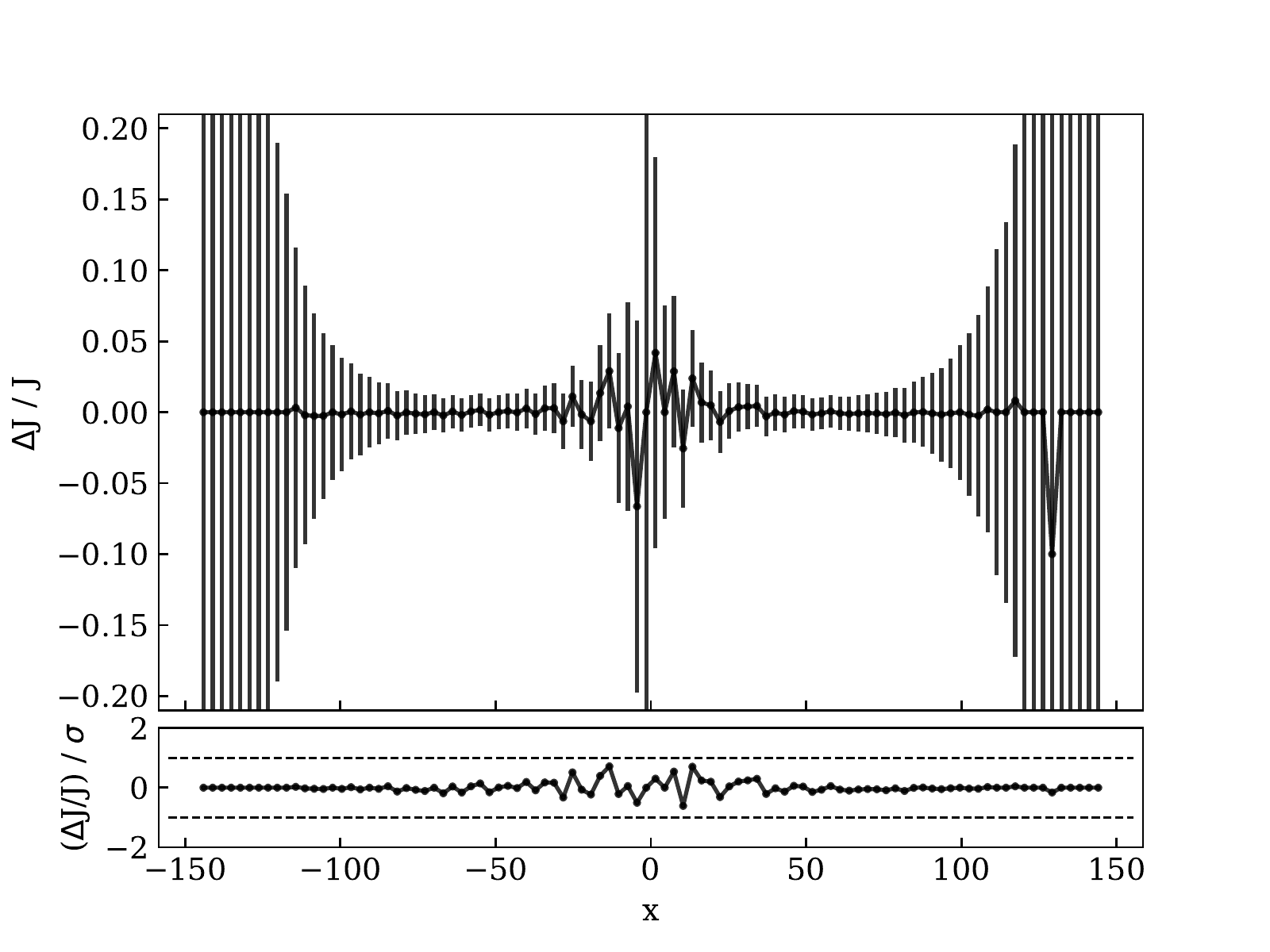}
  \caption{\textsc{rascas} simulations of the transfer of
    Lyman$-\alpha$ photons ($x_{\rm in}=0$) emitted at the origin of a
    uniform plane-parallel slab with $\toto = 10^7$ and $T =
    10$~K. \textit{Top}: Relative difference
    between the emerging spectra computed with the approximation from
    \citet{tasitsiomi_ly_2006} and that computed with the
    approximation from \citet{humlicek_optimized_1982}.
    \textit{Bottom}: As
    in the top panel, but in units of the Poisson error
    (i.e. the variance due to the limited number of photons per bin in
    frequency; $\sigma$). The horizontal dashed lines show
    $\pm 1\sigma$.}
\label{fig:VoigtAccumulatedError}
\end{figure}

We discuss in Sect. \ref{sec:voigt} the systematic errors that
arise from various approximations of the Voigt profile. The question arises of 
how these errors accumulate as a photon performs a random walk in
space and frequency. We illustrate this in
Fig. \ref{fig:VoigtAccumulatedError}, where we show the relative
differences between the spectra emerging from a static slab
illuminated by a central monochromatic \lya{} source and computed with
the different approximations of the Voigt profile. In these
experiments the temperature of the gas is set to $T=10$~K
(i.e. $a\approx0.015$) so as to maximise the errors (i.e. in the
domain of values of $a$ where the three approximations have different
levels of accuracy; see Fig.~\ref{fig:voigt_accuracy}), the opacity of
the sphere is $\toto=10^7$, and the number of photon packets is
$10^6$.  In Fig.~\ref{fig:VoigtAccumulatedError}, we show the relative
difference between the emerging spectra computed using the
approximation from \citet{tasitsiomi_ly_2006} and that computed
using the approximation from \citet{humlicek_optimized_1982}. This
tells us that the two approximations give the same solution at the
$1\sigma$ level. We also did the same comparison for the
approximations given by \citet{smith_lyman_2015} and
\citet{humlicek_optimized_1982} and find the same results. This
suggests that despite the errors made on one computation of the Voigt
function for one value of $x$ and $a$ (see Sec.~\ref{sec:voigt}),
errors do not accumulate noticeably in experiments with a large number
of scatterings.

\section{Distributed radiative transfer on adaptive meshes}
\label{sec:meshrt}

In a typical \ras{} run, radiation is propagated in structured media
defined on adaptive meshes, and within a finite volume. From that
viewpoint, Sect.~\ref{sec:cellrt} describes what happens within a
single simulation cell, and the present section explains how we apply
MCRT to a full AMR
simulation.  One of the main requirements of \ras{} is to limit the
memory footprint of the code so that it can be used to process very
large simulation outputs on supercomputers with limited RAM per
core. This is achieved with (1) domain decomposition
(Sect.~\ref{sec:cdd}), (2) a flexible interface to extract physical
quantities from a simulation output (Sect.~\ref{sec:interface}), (3)
efficient indexing (Sect.~\ref{sec:struc}), and (4) a distributed
master-worker scheme with optimal adaptive load-balancing
(Sect.~\ref{sec:mpi}).

The three first points above are dealt with in a pre-processing step
using the stand-alone code {\tt CreateDomDump}. This code relies on
three \ras{} classes\footnote{Although \ras{} is not strictly
  object oriented, it is very much so in spirit, and we use the
  object-oriented nomenclature when it improves clarity.} to manage
domain decomposition, adaptive mesh indexing, and extraction of
physical data from simulation outputs:
\begin{itemize}
\item The {\tt domain} class defines the geometric properties of a
  domain. This class contains public methods which, for example,     check whether a
  point is within a domain or compute the distance of a point to the
  border of the domain (either the smallest distance or the distance
  in a given direction). \ras{} implements it for a handful of simple
  domain shapes, namely spheres, shells, cubes, and slabs\footnote{The
    slab is literally a slice of the simulation box defined only by a
    thickness in one dimension, and infinite in the two other
    dimensions thanks to periodic boundary conditions}, which are
  defined by a few parameters (their central positions and their size
  or extent).
\item The {\tt gas\_composition} class defines the mixture through
  which radiation propagates (e.g. \ion{H}{i}, deuterium, and dust). It
  implements the conversion of simulation outputs into physical
  quantities useful for the target RT experiment. This class also
  manages the interactions of photons with matter. In particular, it
  contains public methods which return the optical depth along a path
  and which perform scattering events.
\item The {\tt mesh} class handles cells and their indices. The {\tt
    mesh} class is derived from the {\tt domain} class (i.e. a {\tt
    mesh} object is always defined within a domain). A mesh object is
  built from the collection of (leaf-)cells within its domain. It
  contains private methods which recompute all necessary indices (see
  below), and two main public methods which  efficiently  answer two
  questions:  In which leaf cell  is a point?   Which
  neighbouring leaf cell will a photon enter when leaving its current
  cell with a given direction of propagation? The {\tt mesh} class is
  also derived from the {\tt gas\_composition} class, so that it can
  collect and use relevant physical information concerning all leaf
  cells.
\end{itemize}

We  chose to keep this pre-processing step independent of the
MCRT step, in the same spirit as for the casting of photons (see
Sect.~\ref{sec:sampling}). There are  three main reasons. First,  this
step is not CPU expensive;  on the contrary, it is I/O 
expensive and possibly requires a lot of memory. Second,  it may be
necessary to run this code iteratively in order to adjust the free
parameters, for instance the parameters of the domain decomposition.
Last but not least, it uses the simulation data, so it is best
to run this code where the simulation data is stored and then to transfer
the relatively light meshes to the computer where MCRT will be computed.

\subsection{Domain decomposition}
\label{sec:cdd}

\begin{figure}
  \centering
  \includegraphics[width=8cm]{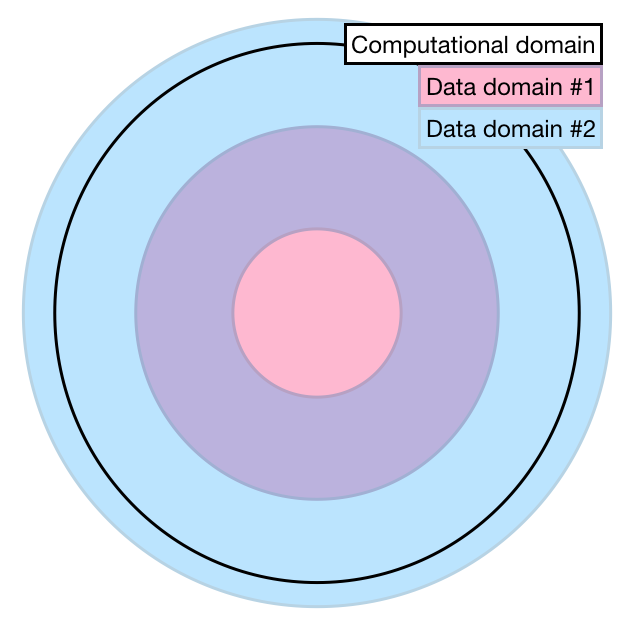}
  \vspace{0.1cm}
  \caption{Example of a domain decomposition with two shell data
    domains (pink and blue overlapping areas) covering a spherical
    computational domain (solid black circle). A photon packet emitted
    from a central source would first propagate through   
    data domain 1 (pink), then be transferred to the data domain 2 (blue),
    and eventually escape the computational domain. The large 
    overlap between domains 1 and 2 (purple) ensures that the photon is
    well within domain 2 before being transferred, which reduces
    possible communication overheads due to photons oscillating between
    domains. }
  \label{fig:decomposition}
\end{figure}

The first domain to define is the computational domain. This is
a unique domain which delimits the volume in which the MCRT is
done. Photon packets are emitted within this domain and their
propagation stops when they reach its border (unless they were
absorbed before). This domain has no data directly associated with it.

We then need to cover this computational domain with a series of
data domains.   Each of these domains defines a mesh object which
contains the physical information of all leaf cells within it. The
propagation of photon packets through the computational domain is done
iteratively through the data domains. When a photon leaves a data
domain, it enters the next one, and so on until it leaves the
computational domain or is absorbed.

A simple example of domain decomposition is shown in
Fig.~\ref{fig:decomposition}, where two data domains (blue and pink
shells) are used to cover the computational domain (sphere outlined
with the black circle). The data domains should overlap significantly
to minimise photon packets bouncing back and forth between data
domains. They should extend slightly beyond the computational domain
in order to make sure that all cells intercepted by the computational
domain are included in at least one data domain. Using small data
  domains has two advantages. The first   is that the propagation
through the mesh should be computationally efficient because it is
compact in memory. The second  is that each mesh has a controlled
and limited size in memory, allowing us to post-process any simulation
whatever the RAM of the machine.

We provide a stand-alone code {\tt CreateDomDump}, which can be used to
perform domain decomposition for some pre-defined typical
geometries. For instance, {\tt CreateDomDump} can produce data domains
as a series of concentric shells centred on one halo
(Fig.~\ref{fig:decomposition}), or as a series of cubes paving a large
simulation box. Importantly, the data domains need not correspond to
the domains that the simulation code (e.g. \textsc{ramses}) has used: {\tt
  CreateDomDump} will collect the leaf
cells belonging to each
data domain by searching in all the simulation domains if
necessary. There are virtually no constraints here except those cited
above: the data domains should cover completely the computational
domain, and large overlaps are best when there are multiple
scatterings.

Finally, as we discuss in Sec. \ref{sec:mpi}, the dynamical
load-balancing scheme of \ras{} makes the computational cost of an
experiment independent of the domain decomposition. In particular,
there is no need to try and define domains that should represent an
equal numerical load: \ras{} dynamically adjusts the number of CPUs
assigned to each domain so that all processors are active at all
times. The main consideration for domain decomposition with \ras{} is
thus mainly the memory footprint: no domain should exceed the
available RAM.

\subsection{Interfaces with simulations}
\label{sec:interface}

Once data domains are defined, {\tt CreateDomDump}   collects leaf
cells within each one and defines their physical properties. This is
done by the class {\tt gas\_composition}. This class needs to be
written explicitly by the user following a generic
template\footnote{The current version of \ras{} provides a generic
  template. It also provides implementations for various mixtures of
  \ion{H}{i}, deuterium, and dust, and for some \ion{Si}{ii},
  \ion{Mg}{ii}, and \ion{Fe}{ii} fluorescent transitions (see
  Table~\ref{table:1}). It is straightforward to build new cases from
  these examples.}. This involves three things. First, it needs to
define attributes that are necessary for RT through a given gas (and
possibly dust) mixture. For example, computing the propagation of
\lya{} photons through hydrogen and dust requires the knowledge of the
density of neutral H, $n_{\rm \hi}$; the velocity of the gas,
$\mathrm{v}_{\rm cell}$; the thermal velocity dispersion of hydrogen
atoms, $\mathrm{v}_{\rm th}$; and the density of dust grains,
$n_{\rm dust}$. Second, the class needs to implement a constructor
that calls external subroutines to define the values of these
attributes for all cells in the domain. In the current version of
\ras{} these external subroutines are built for \ram{}
\citep{teyssier_cosmological_2002} and \ramrt{}
\citep{rosdahl_ramses-rt:_2013}, and packaged in a single {\tt ramses}
module. Such a module may easily be constructed for other simulation
codes, and would simply need, in the above example, to contain
subroutines that will compute $n_{\rm \hi}$, $\mathrm{v}_{\rm cell}$,
$\mathrm{v}_{\rm th}$, and $n_{\rm dust}$. Third, the {\tt
  gas\_composition} class implements public methods that return
useful quantities such as the optical depth along a given distance, or
that perform scattering and/or absorption events. These methods are easily
written, as they also rely on external classes that package
ion-transition and dust properties. The current implementation of
\ras{} provides classes for transitions listed in
Table~\ref{table:1}. They represent a variety of cases (dust,
resonant lines, fluorescent lines) that provide a complete set of
examples for future additions, and that can readily be used in custom
{\tt gas\_composition} implementations.

The second step above is very similar to the {\tt ion\_balance} step
of {\sc Trident} \citep[see][Sect. 2.2]{hummels_trident:_2017}. Here,
we make the slightly different choice to rely on functions that are
very closely connected to the simulation code (\ram{} in our current
implementation) instead of generic functions that, although correct,
may not be fully consistent with the assumptions made in the
simulation. This choice is also motivated by the fact that we are
mostly interested in processing \ramrt{} simulations, which provide
the non-equilibrium \ion{H}{i} density in all cells directly, accounting
for a non-uniform ionising radiation field. This being said, the
modular nature of \ras{} makes it very easy to plug {\tt ion\_balance}
or an equivalent into the {\tt gas\_composition} class instead of our
default {\tt ramses} module.

\subsection{Mesh (re-)construction}
\label{sec:struc}

The final pre-processing step consists in building indices for the
leaf
cells that fill each domain, so that photon packets can be
transported efficiently across the grid. \ras{} uses a graded octree
structure very similar to that of \ram{}, and which requires that two
neighbour leaf cells never have more than one level of difference.  \ram{}
meshes satisfy this condition by construction, so any sample of leaf
cells from a \ram{} simulation output will be useable directly by
\ras{}. The simplest way to use \ras{} to process
  outputs from other simulation codes (e.g. SPH, block-structured AMR, or
  moving-mesh) would be to convert these outputs to a graded octree
  structure. For grid-based codes, an alternative would be for the user to provide
  two routines which (1) efficiently return the index of cell in which
  a photon packet is located, and (2) efficiently return the neighbour
  cell into which a photon packet is moving. The modularity of \ras{}
  makes the replacement of these core routines relatively
  straightforward.

\begin{figure}
  \centering
  \includegraphics[width=8cm]{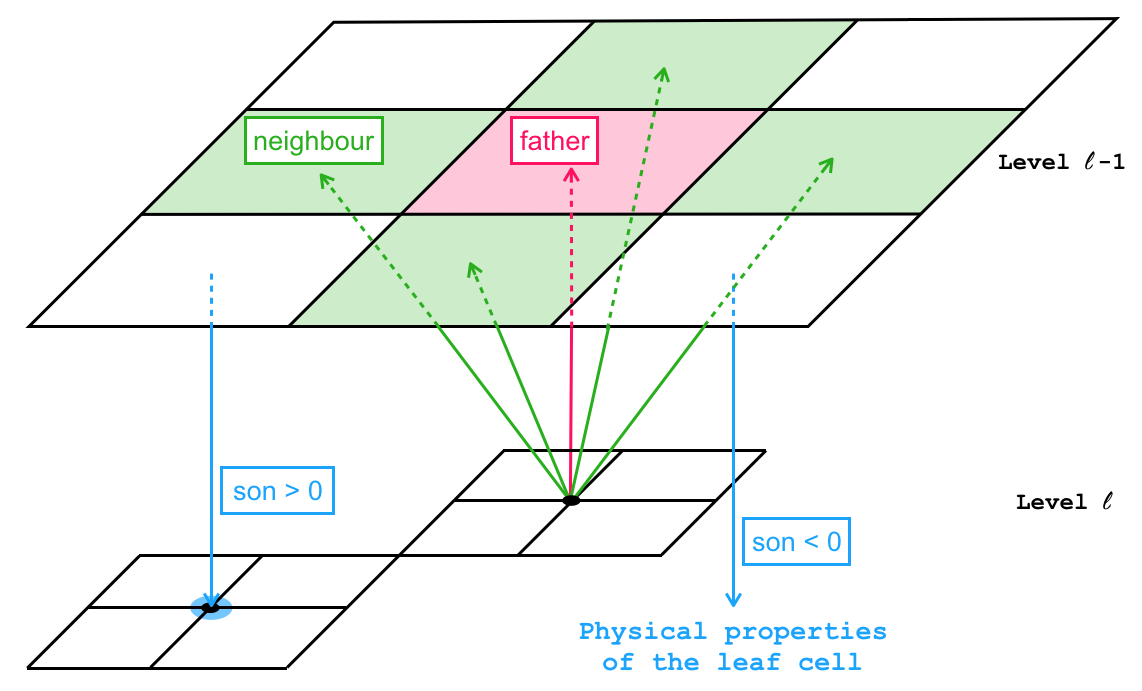}
  \vspace{0.1cm}
  \caption{Two-dimensional illustration of the mesh indexing in
      \ras{}. A level $l$ oct has five pointers (seven in 3D)  to level $l-1$
      cells: the cell containing the oct (father, red arrow) and
      the four neighbours (six in 3D) (green arrows). A cell at level
      $l-1$ has a pointer son either to an oct at level $l$ if
      it is refined (blue arrow, left) or to the
      physical properties describing of the cell if it is a leaf
cell
      (i.e. it is not refined;  blue arrow, right). See text
      for a comprehensive description.}
  \label{fig:amr_struct}
\end{figure}

We store the physical properties of the collection of leaf cells in a
given domain in flat arrays defined by the {\tt gas\_composition}
class. These arrays are relatively small as they are
reduced to the minimum number of physical properties necessary for the
RT computation, and to the number of leaf cells which may be made
arbitrarily small with an adapted domain decomposition strategy. We
also define an AMR tree structure, which allows us to efficiently
locate a leaf cell containing a photon packet or the neighbour leaf
cell into which a photon packet is moving. This AMR structure is
borrowed from \ram,{} and is illustrated in Fig.~\ref{fig:amr_struct}. It
consists of three arrays. The father array gives for each oct at level $l$ the index of the cell at level $l-1$ which
contains that oct and its eight level $l$ cells. The neighbour array gives for each oct at level $l$ the index of the six
cells at level $l-1$ which are neighbours of the father 
cell. These two arrays are relatively small as their size is given by
the number of octs in the domain. The third array, son, is
larger and contains one integer per cell. When the value of son
is positive, it is the index of the oct contained in the cell
(the inverse of the father link). This is the case when the cell
is refined (i.e. when it is not a leaf cell). Such cells are stored solely
for search purposes and have no associated physical information. When
the value of son is negative, which is the case for leaf cells,
then the absolute value of son gives the index of that leaf cell
in the arrays which contain the physical properties of all leaf
cells. The correspondence between the physical-property arrays and
the AMR-tree-structure arrays is thus established with no further
variables. The links between the cells of an oct and that oct are
defined implicitly through the same indexing convention as in \ram{}
and also need no further pointers.

It should be  noted that we build the mesh and indices with depth-first
ordering. This is generally  efficient for neighbour searches
using the AMR tree, which is important for RT applications. The {\tt
  mesh} class features one such routine, which we use abundantly to
locate the cells into which photons are entering after they leave their
current cell. This neighbour search works as follows. First, we check
whether the escaping photon remains within the father cell (at level
$l-1$) of its current cell (at level $l$). This represents 50\% of
cases. If so, we walk the AMR tree from there  to
find the leaf cell into which it is moving (at most two levels). Second, if the photon does
not remain in that father cell, there are three neighbour cells (at level
$l-1$) into which it may go, which are on the three faces of the current
cell that point outside its father cell. There is again a 50\% chance
that the photon will enter any of these three
cells. We check these cells one by one, and in the case of success, walk
the AMR tree from there. Third, it may happen that the two previous
searches are unsuccessful because, although very rarely, photons escape
diagonally (across vertices or corners of cells). In such a case,
testing the many neighbours to determine where photon is might be
relatively expensive. Instead, we search for the target leaf
cell by
walking the AMR tree from the root (level $l=0$).

\subsection{Parallelisation scheme}
\label{sec:mpi}

\ras{} also includes a parallelisation scheme using a master-worker
scheme. We make use of the Message Passing Interface (MPI)
library to communicate between the master and the workers. This is
done by defining an MPI-type photon packet. One MPI thread is the
master task and all other MPI threads are worker tasks. The master
coordinates work with two key ingredients: a photon-packet queue for
each data domain, and an adaptive mapping of workers to data domains
(see Sect.~\ref{sec:cdd}).

In practice, our algorithm works as follows. At initialisation, the
master first builds data-domain queues with all photon packets
depending on their positions: a photon packet is assigned to the
data domain that contains it, and if more than one contain it,
  to the one within which the photon is furthest away from
any border. In other words, photon packets are assigned to
data domains in which they will travel the most before moving out, thus
minimising data domain changes. Second, the master balances the
computational load through domain--worker mapping: the number of
workers attributed to each data domain is proportional to the number
of photon packets in each data-domain queue. When this is done, the
master sends a first series of bundles of photon packets, one to each
worker, and waits for any worker to send back its processed
bundle. Since the computational time to process a bundle of photon
packets is not constant, the communications between the master and the
workers are asynchronous.

When the master receives a bundle of photon packets back from a
worker, it first determines what to do with each photon packet in the
bundle. There are three possibilities. First the photon packet may
have escaped the computational domain. In that case the RT is done and
the master updates the final properties of this photon packet in order to save it to disc at a later time. Second, the photon packet may have left the
data domain of the worker, while remaining within the computational
domain. In this case, the master finds the new data domain to which
the photon packet belongs, and appends it to the corresponding
queue. Third, in the case of RT with a dust component, a photon packet may
have been absorbed by a dust grain. In this case, as in the first
case, the RT is done, and the master only updates the final properties
of this photon packet before saving it to disc. The photon packet type
carries a {\tt status} flag which indicates  why the computation
ended for each photon packet.

After having processed a bundle of photon packets, the master checks
that the CPU mapping is still adequate, meaning  that the worker that
sent back the current bundle is assigned to a data domain that has
more than $N_{\rm bundle}$ photon packets in its queue, where
$N_{\rm bundle}$ is the number of photon packets in a bundle. If this
is the case, the worker keeps its current data domain. On the
contrary, if the worker's data domain has an empty queue, the worker
is assigned to a new data domain. In this case, we choose the
data domain $i$ with the highest ratio
$\delta(i) = (N_{\rm packet}(i)/N_{\rm bundle})/N_{\rm worker}(i)$, where
$N_{\rm packet}(i)$ is the number of photon packets in the queue of
data domain $i$, and $N_{\rm worker}(i)$ is the number of workers
associated to the data domain $i$. Finally, the master prepares a new
bundle of photon packets from the queue of data domain $i$ and sends
it to the worker. The sending phase is always done in two steps. First,
the master sends a data domain ID to the worker. The worker receiving
the data domain ID checks whether it is the data domain already loaded in
memory, and if not,  loads it. Second, the master sends the bundle
of photon packets, which the worker receives and processes.

This parallelisation scheme has the great advantage of implementing an
optimal adaptive load-balancing. Although the queues of different
data domains are not balanced at all times, the workers are always at
work and the number of data domain changes is minimal. This latter
feature is important as it limits a small overhead due to I/O.   Also,
by construction, the computational load of the master is low, and it
spends most of its time waiting for communications from the
workers. We use blocking communications (the \texttt{MPI\_Send} and \texttt{MPI\_Recv}
functions), but since the communications are asynchronous and since
there is no synchronisation during the run, the time spent by each
worker  waiting to send or receive data is extremely limited, and
so workers spend most of their time in computing.

Finally, it is also worth mentioning that thanks to this
parallelisation scheme, we implement domain decomposition at no CPU
cost, which allows us to minimise the memory footprint of the
code. Each worker has only the mesh and gas composition data for its
data domain, whereas the master has only minimal information (shape
and extent) for all domains and photon-packet lists. In practice,
data domains may be very small (easily less than 100MB) making \ras{}
usable on any architecture to process arbitrarily large simulations.

\subsection{Basic test and error budget}

Here, we perform some basic tests to demonstrate the precision and
efficiency of \ras{} when running RT simulations on full mesh. The
test experiment is the propagation of photon packets through a
homogeneous and static sphere. In order to assess the precision of the
code, we impose that scattering events do not change the propagation
direction of photons, so that the distance travelled  by all photon
packets should be exactly the radius of the sphere, at the precision
of the code.

In a first series of experiments, our goal was to measure the numerical
error due to our treatment of scattering events only. For this, we used
a particular set-up in which the sphere is fully included within one
simulation cell. We set the density of the medium so that the
optical depth from centre to border, $\toto$, takes values $10^{-10}$,
$10$, and $10^6$. We cast $N_{p} = 10^6$ photons (only $10^3$ for the
experiment with $\toto=10^6$)  from the centre of the sphere
in random directions, and integrated their travel distances.  We then
compared these distances to the theoretical distance. For the run with
$\toto = 10^{-10}$, there is no scattering event, and the integrated
distance is exactly equal to the theoretical distance. In other
words, the relative error is zero for all photons.
For $\toto = 10$, almost all photon packets have scattered a few
times, and there is a tiny difference between the integrated distance
and the theoretical one. These errors are of the order of $10^{-15}$,
 the median is zero and $98\%$ of the photon packets have an error
between $-9\times10^{-16}$ and $9\times10^{-16}$. This error is comparable
to the numerical precision $\sim 2 \times 10^{-16}$ of floating point in
double precision in Fortran\footnote{This value is given by the
  intrinsic function {\tt epsilon}, which returns the smallest number
  $E$ such that $1 + E > 1$.}.
For $\toto = 10^6$, all photon packets have scattered many times
(almost one million  scattering events), and we observe that the
median value of the relative errors remains equal to zero, but the
dispersion increases. For $98\%$ of the photon packets, the relative
error is between $-2.5\times10^{-13}$ and $2.5\times10^{-13}$.
This error, which increases with the number of scattering events and 
with $\toto$, as expected, is simply due to the inaccuracy in
computing a distance as the sum of ever smaller segments. We
find that the amplitude of this error   remains very low, even in relatively
extreme cases, which validates our implementation.

In a second series of experiments, our goal was  to measure the numerical
error associated with the propagation of photon packets through a
grid. Here, we repeated the same experiments as above, but this time
with the sphere filling up a regular mesh of $256^3$ elements. This
time, for each of the three values of $\toto$, we find that
the median value of the relative errors is the same for the three
experiments, and is equal to $\sim 6.7\times10^{-14}$. The dispersion of
these relative errors also remains   approximately the same:   a relative error between $\sim
-3\times10^{-14}$ and $\sim 2\times10^{-13}$ for
$98\%$ of the photon packets. 
This error is mostly due to repeated changes of coordinates at each
cell crossing, where the position of a photon is converted from a
position in the frame of the current cell, in cell-size units, to a
position in the global simulation frame, in simulation-size units, and
back to a cell position. Empirically we find that this error is
roughly proportional to three times the numerical precision (in double
precision) times the number of cells crossed. Again, this shows that the
amplitude of the errors is very low, and is well below other
uncertainties inherent to the numerical implementation of radiative
transfer physics (see Sect.~\ref{sec:mcrt}).

We note that we find the same emergent spectrum for the case of the
static uniform \ion{H}{i} sphere embedded within one cell and the case
of the sphere distributed on a $256^3$ mesh (see
Fig.~\ref{sphere}). This demonstrates that the relative errors
discussed above are indeed very small with respect to the target
result.

Finally, we compare the computational time for all the tests discussed
above. We find that when there is no scattering (i.e. for very low
values of $\toto$) the overhead of the mesh is important and the
computational time may be increased by a factor greater than 10. This
is expected because there is basically nothing to compute in these
cases, and the small overhead due to cell changes will be relatively
important. For higher $\toto$, the RT computation through the
mesh is only $\sim 10\%$ slower than without the mesh.

\subsection{Scaling of the code}

\begin{figure}  
  \centering
  \includegraphics[width=9.cm]{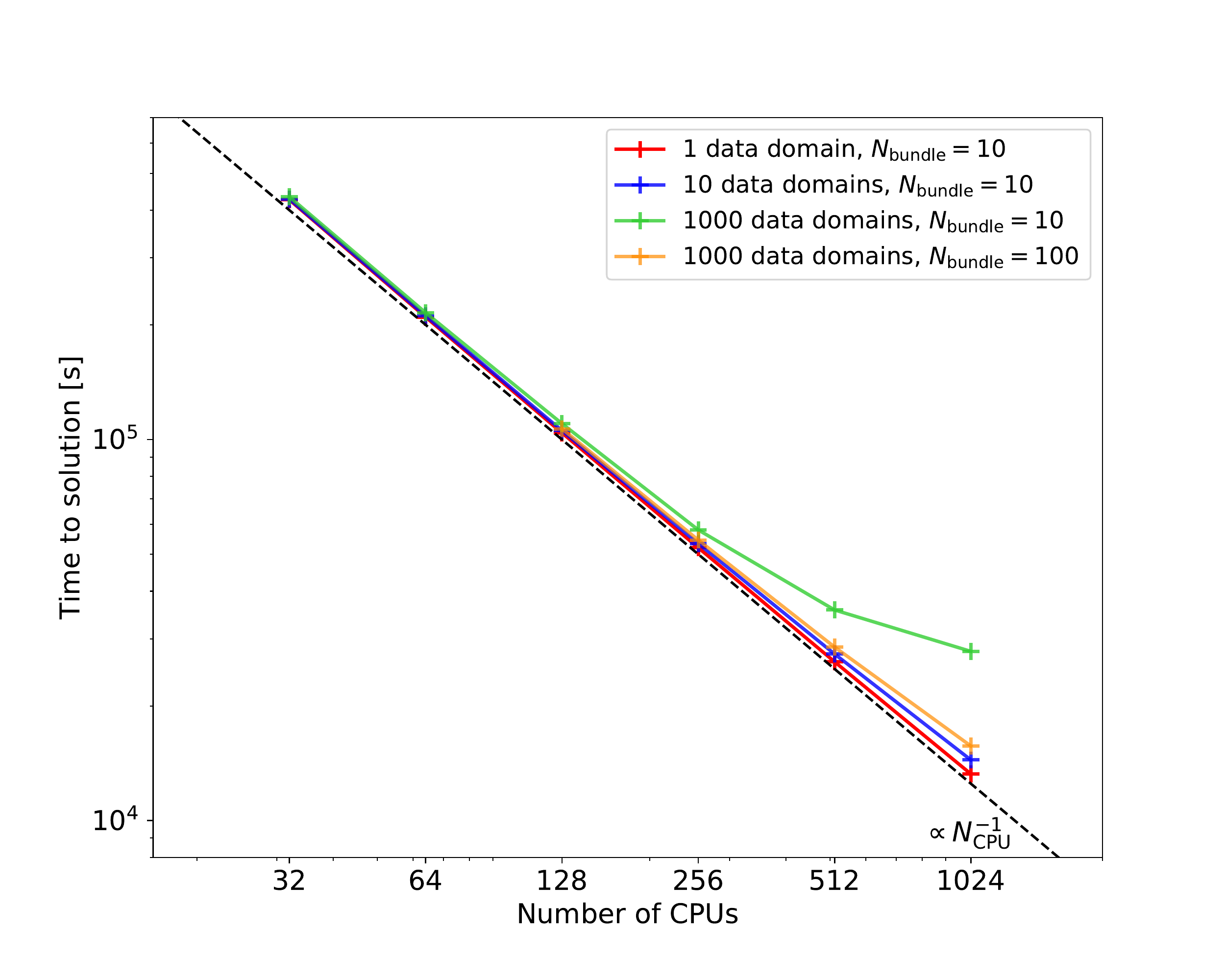}
  \caption{Scaling test of \textsc{rascas} based on an
      idealised galactic disc simulation: total elapsed time for a
      realistic MCRT experiment as a function of the number of CPUs
      used for the run. Red symbols show the results for the series
      where the computational domain is decomposed into one single
      data domain, whereas the blue crosses (respectively green
      crosses) are for the case where the computational domain is
      decomposed into 10 (resp. 1000) data domains. These three
      simulations were   run with $N_{\rm bundle}=10$. Orange crosses
again      show   the case where the computational domain is decomposed
      into 1000 data domains, but this time using $N_{\rm bundle}=100$. To
      guide the eye, the dashed line shows the ideal case of relation
      $\propto N_{\rm CPU}^{-1}$.}
  \label{fig:scaling}
\end{figure}

In this section, we discuss the scaling of the code with the number of
CPUs. We used \ras{} in a realistic set-up by running the
following experiment. We propagated $10^6$ monochromatic \lya{} photon
packets emitted in star-forming regions through the ISM and CGM of an
idealised disc galaxy simulation. The gas is composed of
\ion{H}{i}, \ion{D}{}, and dust. Photon packets were propagated until
either they escape the virial radius of the dark matter halo or they
are absorbed by dust.

We repeated the same experiment for various numbers of CPUs ranging from
32 to 1024. The results are shown in Fig.~\ref{fig:scaling}  where the almost perfect scaling of \ras{} can be
seen. This series of runs was done with one single
data domain. In principle, this is the most favourable configuration,
as it limits both the amount of communication and the number of times
CPUs have to load or unload domains from the disc. However, as discussed in
Sec. \ref{sec:mpi}, in most cases we expect the efficiency of \ras{} to be independent of the domain decomposition. In order to verify
this, we   also performed the same test with the computational
domain decomposed in 10 (blue symbols) and 1000 (green symbols)
data domains. As can be seen in Fig.~\ref{fig:scaling}, the total
elapsed times for 1 and 10 data domains are very close, confirming our
expectations. However, for 1000 data domains the code starts to not
scale perfectly for large numbers of CPUs. This is probably due to the
very small size of the data domains, which results in photon packets
moving from one domain to another after relatively light
calculations. This raises the overhead due to communications to a
noticeable level which decreases the overall performance.

To confirm this, we performed a profiling analysis of the three runs
with $N_{\rm CPU}=512$.  With one data domain, we find that the
$N_{\rm CPU}-1$ workers spend most of their time  computing
($>98\%$), whereas the master spends less than $3\%$  computing (i.e. it spends $97\%$ of its time   waiting for messages from the
workers). It is thus highly available, and the code scales perfectly.

With ten data domains, the number of communications increases by a
factor of three compared to the previous case. Then, the
$N_{\rm CPU}-1$ workers still spend most of their time  computing
($\sim95\%$), but start to spend $\sim5\%$ of their time  receiving
messages from the master. Since we use blocking communications, and
since the sending back of messages to the master of the same amount of
data does not cost any time, the workers basically wait for messages
from the master  because the master has more work to do in
managing queues, load-balancing, and sending and receiving messages from
the workers. It thus  spends $\sim9\%$ in computing, decreasing
 its availability slightly.

With 1000 data domains, the number of communications increases by a
factor of nine compared to the case with 10 data domains. The master
now spends most of its time  computing
($\sim60\%$), and the workers wait a lot. Their computing times span a broad range from $46\%$ to $92\%$ of their time, with half of
them spending less than $75\%$ in computing time. In such extreme
cases, the number (and the frequency) of communications is so high
that the availability of the master drops completely, resulting in a
situation where workers spend from $8\%$ to $54\%$ of their time
waiting for messages instead of working.

Fortunately, there is the free parameter $N_{\rm bundle}$, that   controls the number of communications. All the computations
presented so far have been run with $N_{\rm bundle} =10$.  A profiling
analysis of the run with $N_{\rm CPU}=512$ and 1000 data domains, but
this time using $N_{\rm bundle} =100$ shows that by increasing the
size of the bundle of photon packets sent to each worker, the number
of communications drops significantly (by a factor $\sim4)$. Thus, the
master has less work and spends $22\%$ of this time  computing,
increasing significantly its availability. As a consequence, workers
wait less and compute more (between $85\%$ and $99\%$).  To confirm
that we can recover very good scaling by adjusting the size of the
bundle of photons, we performed a series of runs with the
computational domain decomposed into 1000 data domains and using
$N_{\rm bundle}=100$ (shown with orange crosses in
Fig.~\ref{fig:scaling}).  The total elapsed time is then comparable to
the case with one or ten data domains.

In conclusion, we   showed that even in the extreme case of 1000
data domains and for large number of CPUs, the overhead by increasing
the number of communications can be drastically reduced by increasing
$N_{\rm bundle}$. This reduces the number of master--worker
communications and then increases the availability of the master.
This is one necessary condition that explains why the code scales so
nicely with the number of CPUs. Thus, \ras{} can be used with no
overhead for an arbitrarily large number of data domains, and it can
thus be used to process arbitrarily large simulations even with very
limited RAM per core.

\section{Example applications}
\label{sec:applications}

In this section we illustrate how \ras{} can be used to construct mock
observations which can be directly compared to true observations. It
is also worth noting that \ras{} has already been used to compute the
radiative pressure due to multiple scattering of \lya{} photons on hydrogen
atoms, and to develop a sub-grid model for early \lya\ feedback in
\citet{kimm_impact_2018}.  Because of its modularity, \ras{} can also
easily be adapted for a number of different applications. For example,
its ray-tracing engine has been used to estimate escape fraction of
ionising photons from simulated high-redshift galaxies in
\citet{trebitsch_escape_2017}, \citet{costa_quenching_2017}, and
\citet{rosdahl_sphinx_2018}.  Other `by-products' have also been
developed to extract and manipulate sub-volumes in large AMR
simulations, and to compute column densities along any arbitrary
lines of sight.

\subsection{\lya{} image of a high-redshift galaxy} 
\label{sec:lyaExample}

\begin{figure}
  \centering
  \includegraphics[width=9.cm]{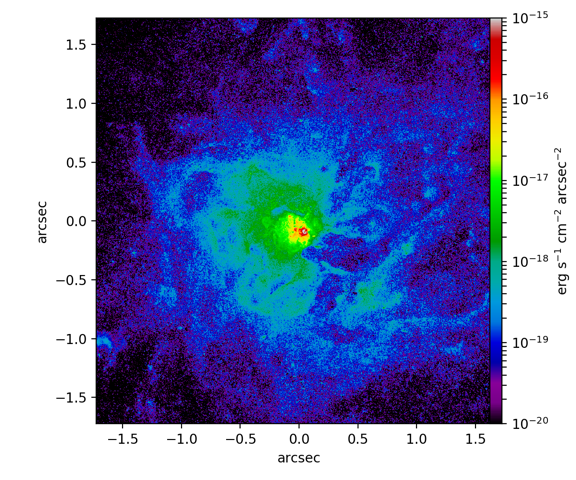}
  \caption{Surface brightness map of \lya{} emission from a
    high-redshift galaxy and its circumgalactic medium. }
  \label{fig:lyamap}
\end{figure}

The main driver for the development of \ras{} is the study of the
\lya{} properties of galaxies in the high-redshift Universe. In
Fig.~\ref{fig:lyamap}, we show a surface brightness map of \lya{}
emission from a simulated galaxy at redshift $\sim4$. The simulated
galaxy has a stellar mass of $\sim 10^9 M_\odot$ and a star formation
rate of $\sim 3 M_\odot/$yr.

The simulation and \ras{} post-processing are fully described in
Blaizot et al., in prep. It is a radiation-hydrodynamic
simulation ran with \textsc{ramses-rt} \citep{rosdahl_ramses-rt:_2013,
  rosdahl_scheme_2015, katz_interpreting_2017, rosdahl_sphinx_2018},
with a spatial resolution of $\sim 15$~pc in the ISM. The sub-grid
models used for star formation and feedback are those presented in
\citet{kimm_towards_2015, kimm_feedback-regulated_2017}, based on the
same calibration as in the SPHINX simulations
\citep{rosdahl_sphinx_2018}.

To construct Fig.~\ref{fig:lyamap}, we  used $6\times 10^6$ photon
packets, emitted from $\sim 10^6$ gas cells within the virial radius
proportionally to their \lya{} luminosities, accounting for 
recombinations and for collisional excitations. We  included dust
following \citet{laursen_ly_2009}, assuming SMC properties. We 
used the peeling algorithm \citep{yusef-zadeh_bipolar_1984,
  zheng_monte_2002} as described in
\citet[][Sect. 9.3]{dijkstra_saas-fee_2017} to collect flux on a
$1000\times1000$ image in a particular direction. In order to
accelerate the computation, we  also implemented the core-skipping
algorithm described in \citet[][Sect. 3.2.4]{smith_lyman_2015}, which
slightly underestimates the effect of dust, but produces a speed-up of a
factor $\sim 1000$.

\subsection{Line transfer in idealised set-ups}
\label{sec:idealsetups}

\begin{figure}
  \vspace{-2.0cm}
  \begin{minipage}{\linewidth}
    \hspace{0.4cm}
    {\includegraphics[width=0.9\linewidth,height=280pt]{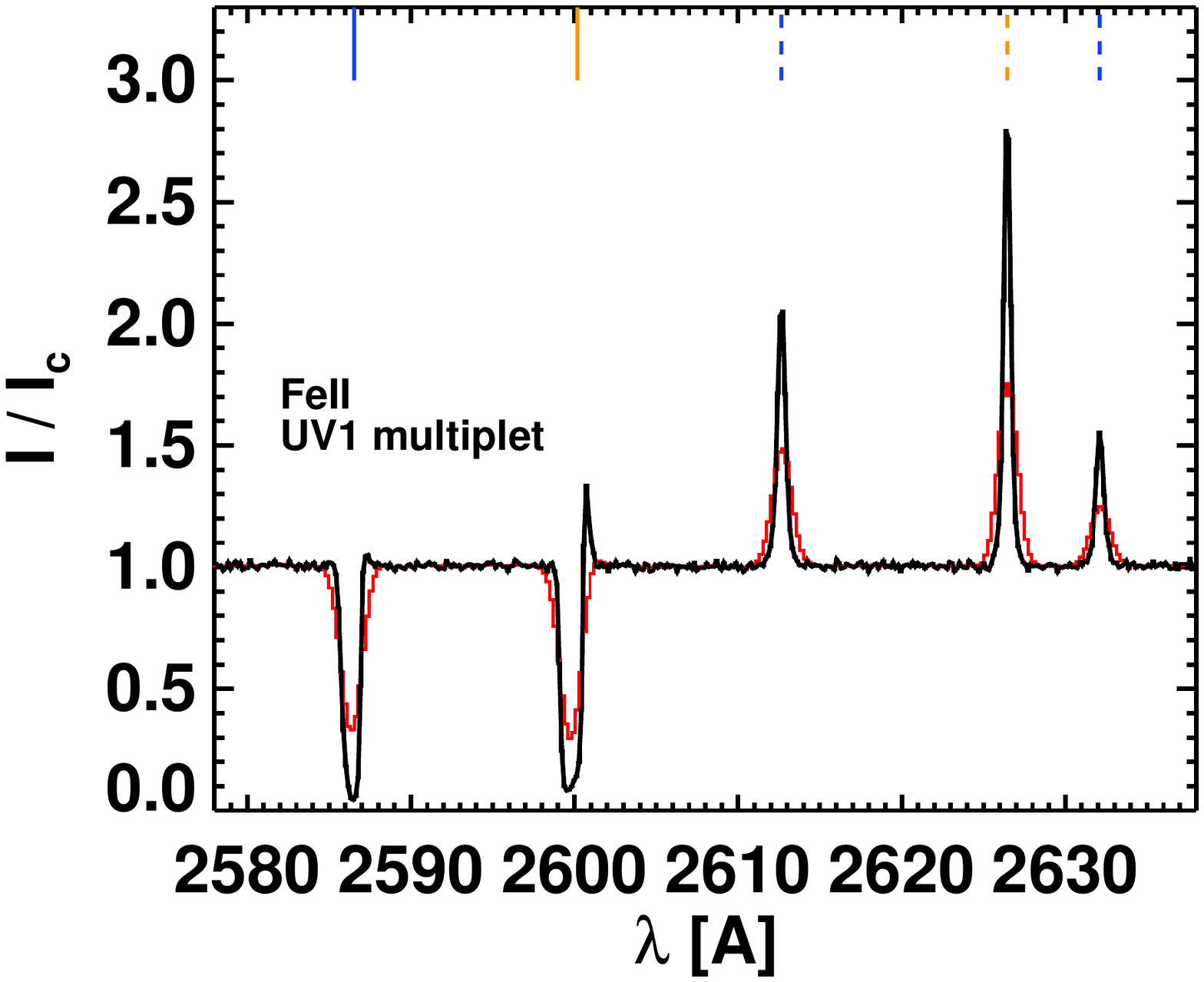}}
    \vspace{-3.8cm}
  \end{minipage}\quad
 \hspace{-0.5cm}
  \begin{minipage}{0.48\linewidth}
    \hspace{0.1cm} 
    \vspace{0.0cm}
    {\includegraphics[width=1.1\linewidth,height=195pt]{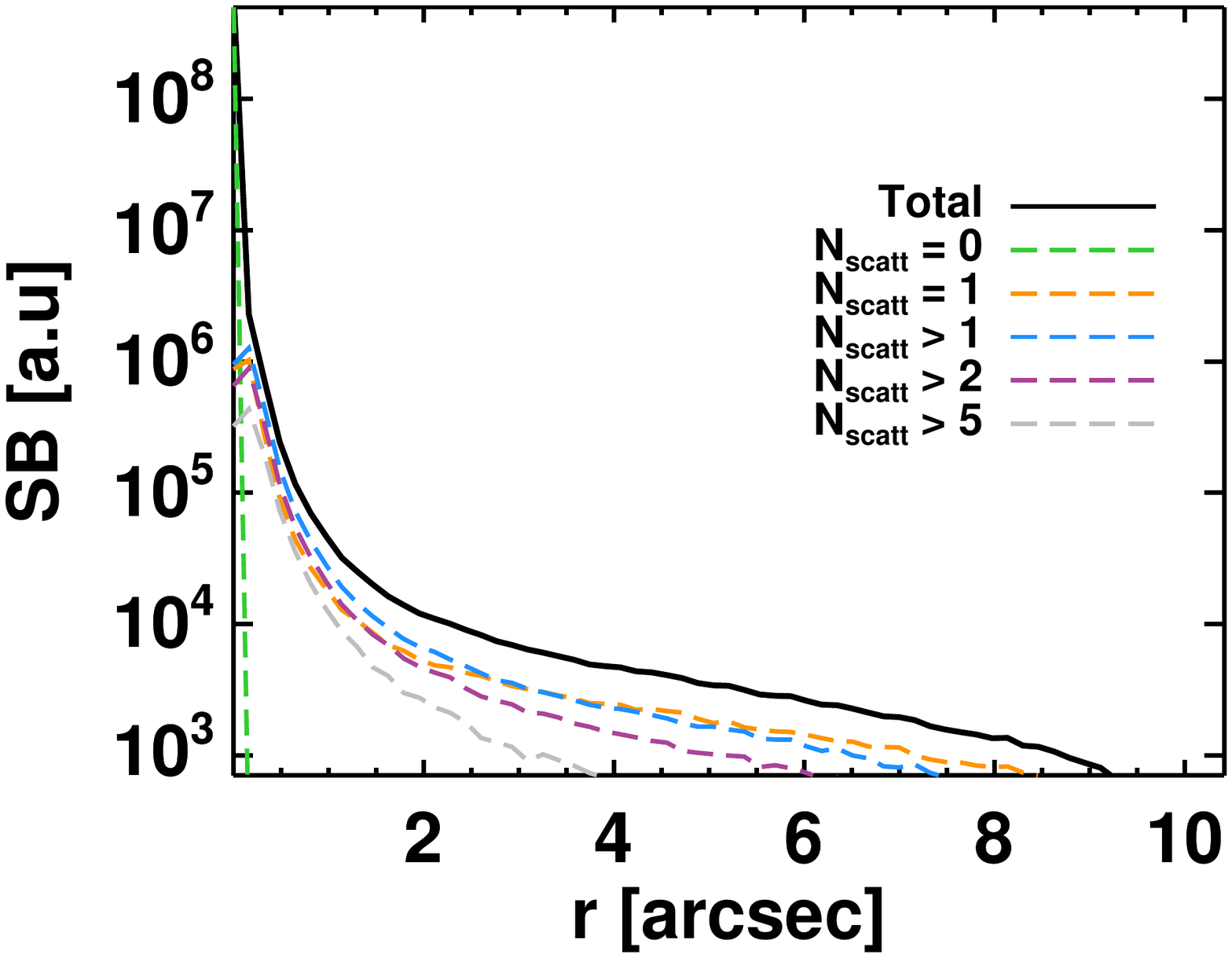}}
    \hspace{-1.2cm}
          \end{minipage}
          \hspace{-0.5cm}
          \begin{minipage}{0.57\linewidth}
           \hspace{0.05cm}
           \vspace{-0.0cm}      
           {\includegraphics[width=1.05\linewidth,height=205pt]{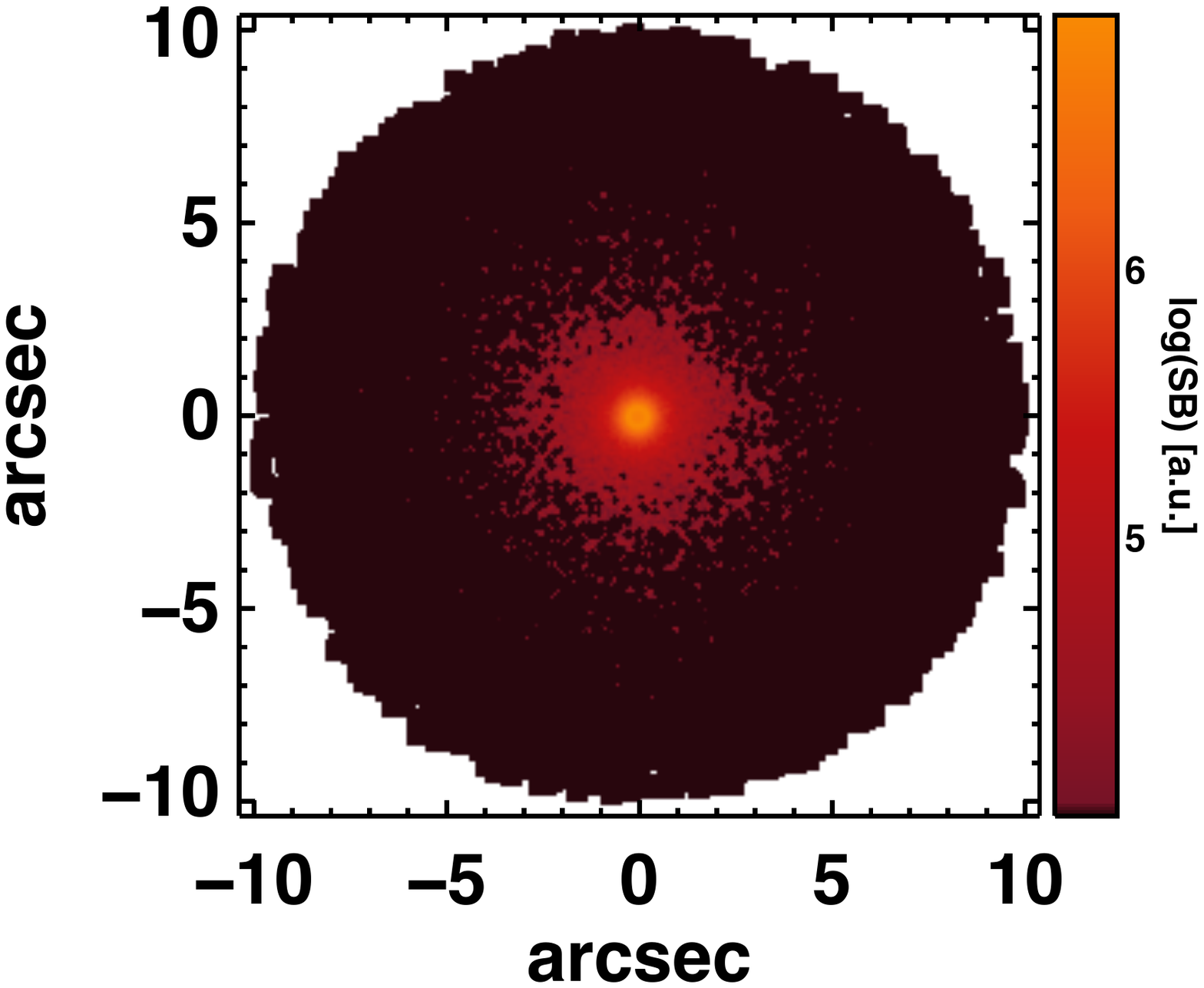}}
  \end{minipage}
  \vspace{-1.6cm}
  \caption{Simulation of the radiative transfer of the \ion{Fe}{ii}
    UV1 multiplet. As  input source we cast a flat UV continuum
    ($\lambda=2570-2640$ \AA) propagated through a galactic wind. In
    our idealised model the gas velocity ($v(r) \propto r$) increases
    from 0 to 100 km s$^{-1}$ and the density profile is described by
    a power law ($\rho(r) \propto r^{-3}$) normalised to an integrated
    \ion{Fe}{ii} column density of $10^{17}$ cm$^{-2}$. The Doppler
    parameter, accounting for thermal turbulent motions, is set to
    $b=30$ km s$^{-1}$. \textit{Top}: Emergent spectrum of the
    \ion{Fe}{ii} UV1 multiplet composed of two absorption features and
    three emission lines (black line).  The vertical solid lines show
    the rest-frame wavelengths of the two resonant transitions
    ($2586, 2600$ \AA), while the dashed lines show the three
    associated fluorescent emission lines ($\approx 2612, 2626,$ and
    $2632$ \AA). The red curve shows the Gaussian smoothed spectrum
    assuming $FWHM=1.2\AA$. \textit{Bottom}: Projected image
    colour-coded by surface brightness (right) and radial surface
    brightness profile (left) at $\lambda=2570-2640$ \AA{} (in
    arbitrary units). The total surface brightness profile is
    represented by the solid black line, and the coloured dashed lines
    correspond to photons that scattered $0, 1, >1, >2$, and $>5$
    times. The image of the extended emission (right panel) has been
    convolved with a Gaussian assuming a PSF full width at
    half maximum of $0.1^{''}$.}
\label{fig:fe2}
\end{figure}

In addition to hydrodynamic simulations, \textsc{rascas} can also be
run on custom idealised models in which users can decide the sampling
and location of the sources as well as the geometry and the properties
of the medium.  These kinds of set-ups are commonly used to guide the
interpretation of observational data (e.g. line profiles) and to test
physical scenarios (e.g. inflows and outflows) based on simplified
assumptions \citep[e.g.][]{ahn_p_2003, verhamme_3d_2008,
  prochaska_simple_2011, laursen_non-enhancement_2013,
  scarlata_semi-analytical_2015}.  As an illustration, we show in
Fig.~\ref{fig:fe2} the output of an RT experiment of a flat UV
continuum ($\lambda=2570-2640$ \AA) in a galactic wind. In this
example, photon packets are emitted at the centre of a spherical
expanding outflow that extends from $r_{\rm min}=1$ to
$r_{\rm max}=20$~kpc. The medium is filled with \ion{Fe}{ii} ions
distributed according to a given velocity profile ($v(r) \propto r$)
and density profile ($\rho(r) \propto r^{-3}$).

The top panel of Fig.~\ref{fig:fe2} depicts the resulting spectrum
in which the various features of the UV1 multiplet of \ion{Fe}{ii} are
visible (black line). Strong absorption lines arise at the location of
the \ion{Fe}{ii} $\lambda\lambda$2586, 2600 resonant transitions. We
note that these two lines are slightly blueshifted due to the bulk
motion of the gas, whereas the broadening of the absorption is
primarily driven by the velocity dispersion and the column density of
the gas. We see that scattering in the outflowing medium gives rise to
an emission redward of the $\lambda\lambda2586$, $2600$ resonances. The
other three emission lines at $\lambda \approx 2612, 2626,$ and $2632$
\AA{} correspond to the fluorescent channels associated with the
resonant transitions (see Sect.~\ref{subsec:scatt_event}). In our
example the convolution of the spectrum with a Gaussian line spread
function (overlaid in red), mimicking a typical instrumental spectral
smoothing, erases the P Cygni-like features close to the resonances
which then appear as pure absorption lines.

Photons emitted by a central source can also resonantly scatter in
physical space giving rise to spatially extended emission that traces
the surrounding gas. This is shown in the bottom right panel, which
represents a mock projected map of the emission of the radiation
emerging from the outflow. In the bottom left panel of
Fig.~\ref{fig:fe2} we plot the surface brightness profile of the
emission at $\lambda=2570-2640$ \AA{} (black curve). The emission is
strongly peaked at $r=0$ because a large fraction of the photons are
far away from the resonance and therefore escape directly without
scattering (in green). Depending on the opacity of the medium, photons
undergoing more than one scattering can contribute significantly to
the total surface brightness profile, as is the case in our example
in Fig.~\ref{fig:fe2} (coloured dashed lines). This highlights the
importance of taking multiple scatterings into account while modelling
the radiative transfer of resonant lines, and therefore the need of
performing MC numerical simulations, in order to construct mock
observables. Finally, we note that $5 \times10^6$ photon packets
were cast from the source and propagated through the wind until escape
of the medium in this simulation, and that it ran on a single core in
a few minutes only.

\subsection{RGB maps of stars with dust from cosmological simulations}

\begin{figure*}
  \centering

  \includegraphics[width=5.75cm]{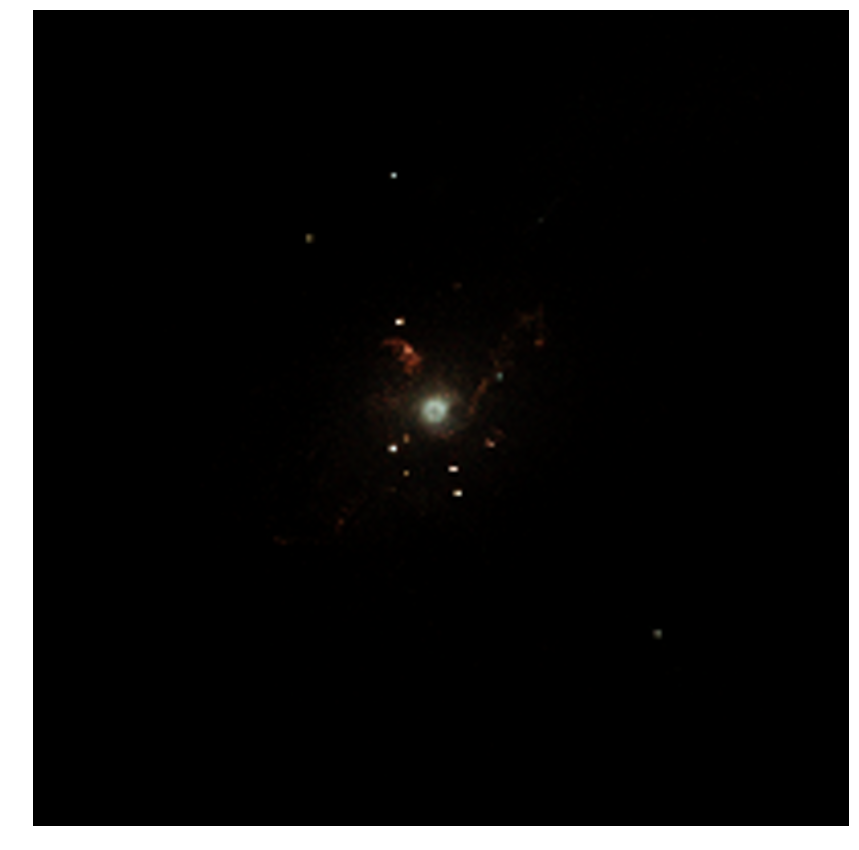}
  \includegraphics[width=5.75cm]{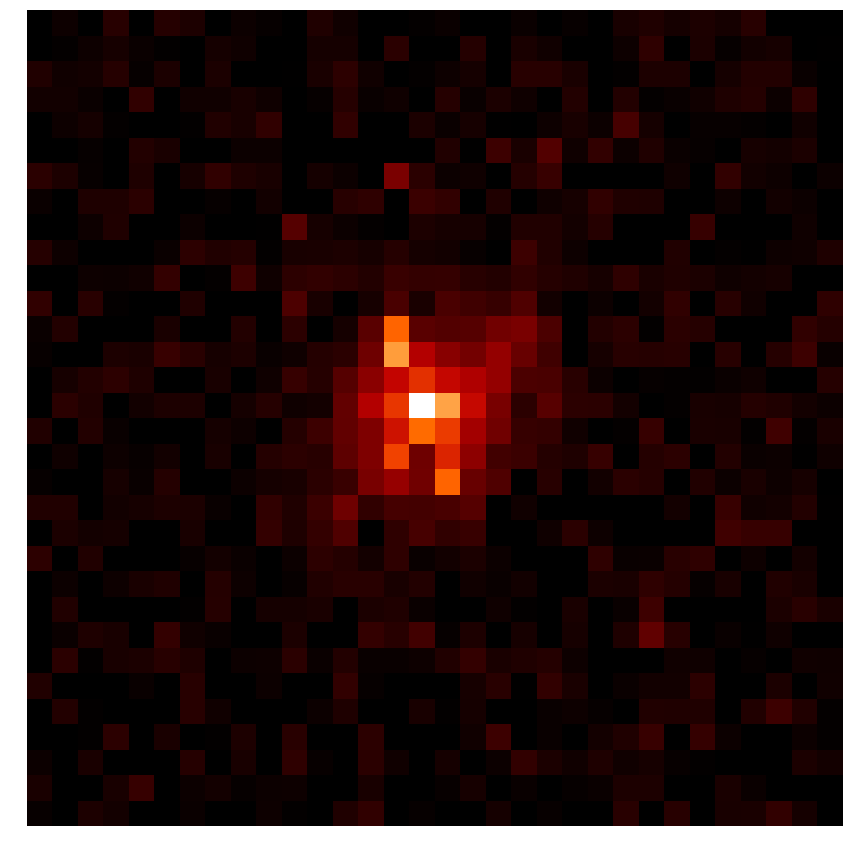}
  \includegraphics[width=5.75cm]{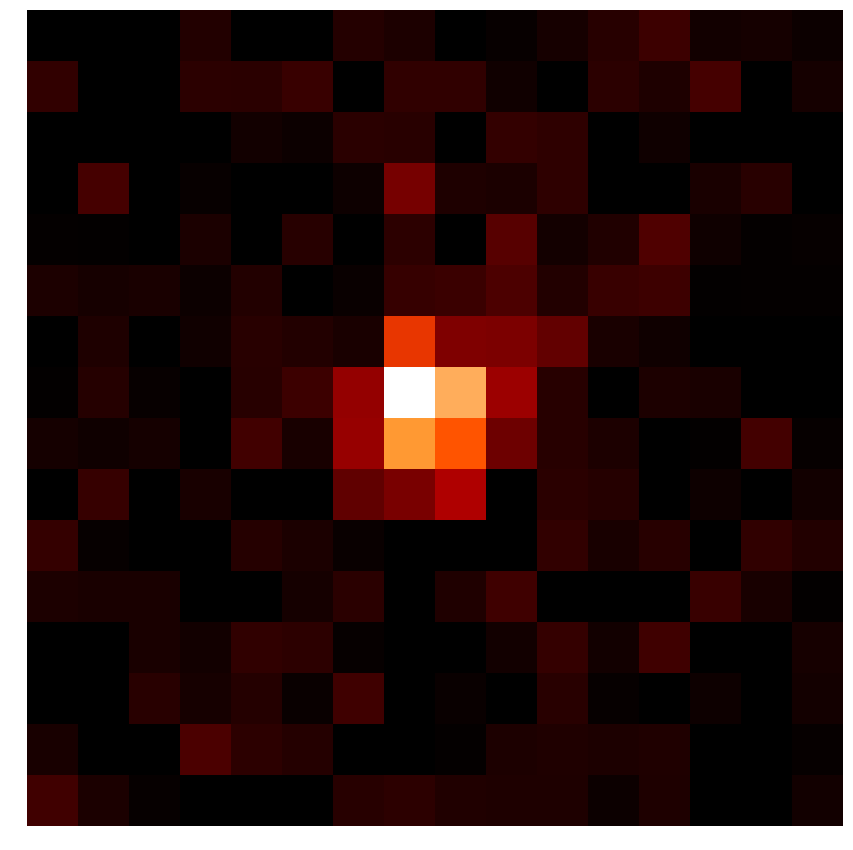}

  \includegraphics[width=5.75cm]{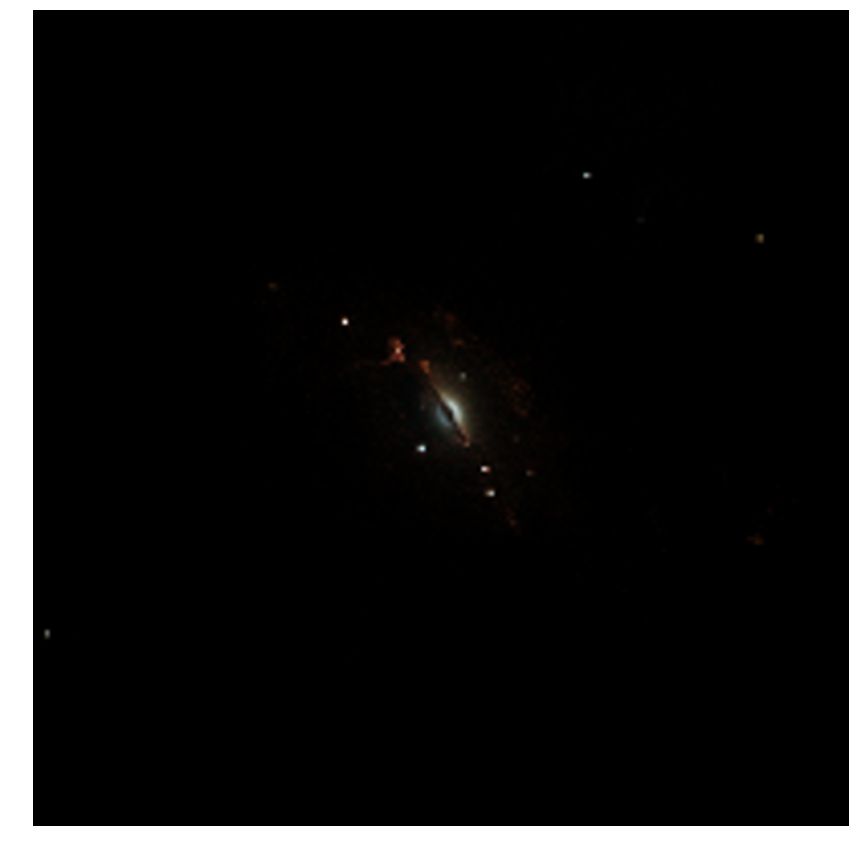}
  \includegraphics[width=5.75cm]{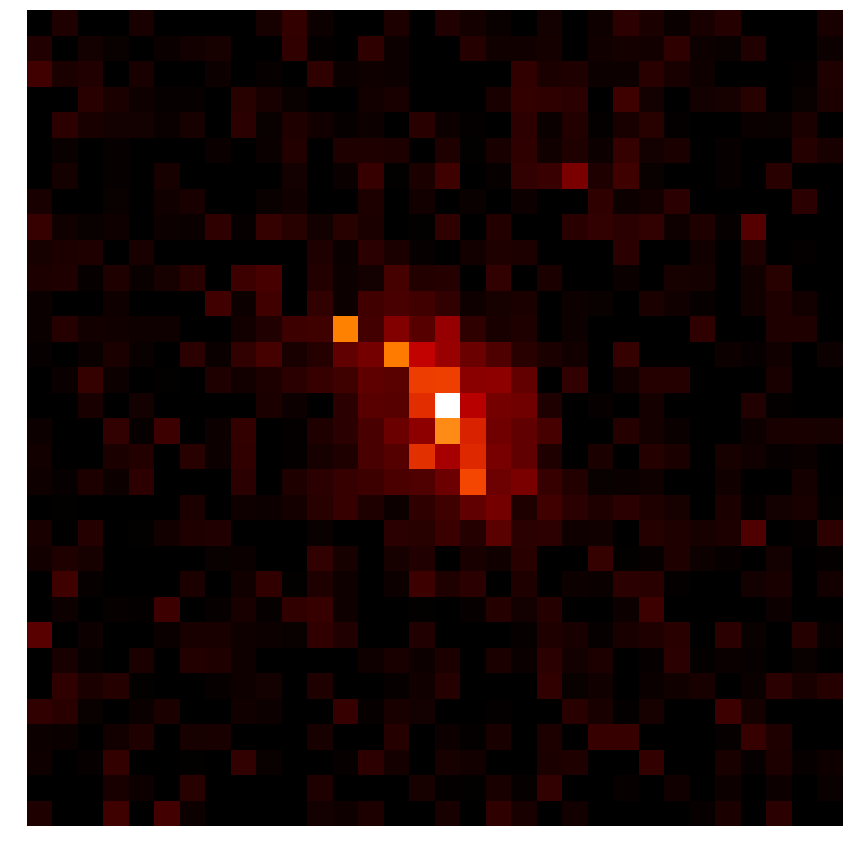}
  \includegraphics[width=5.75cm]{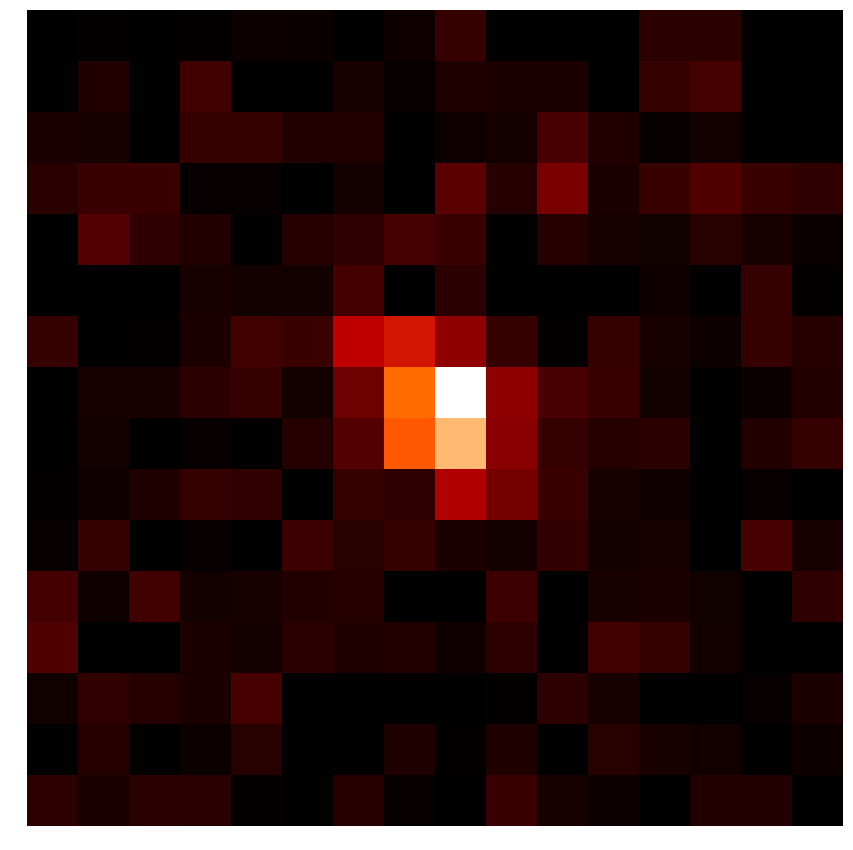}
   \vspace{0.1cm}
   \caption{Mock JWST/NIRCam images. The top and bottom panels show
     two different projections of the same object. \textit{Left:} Pseudo-colour
     RGB image with R=F444W, G=F277W, B=F150W. \textit{Middle:} Mock image in
     the F150W filter at the resolution of JWST/NIRCam in the short-wavelength channel. The noise level corresponds to 1Ms exposure
     time. Colour-coding indicates the flux level in units of
     erg/s/cm$^2$. \textit{Right:} Same, but for the F444W filter
     in the long-wavelength channel. The field of view is the same for
     each panel ($\sim 1$ arcsec $\sim 5.8$~kpc).}
  \label{fig:rgb}
\end{figure*}

\textsc{rascas} can also easily be used to compute radiative transfer
of continuum light in the presence of dust, and then to produce
realistic mock images.  As an illustration, we show in
Fig.~\ref{fig:rgb} an example of \textit{James Webb} Space Telescope
(JWST) NIRCam images. We note,
contrary to dedicated dust MCRT codes
\citep[e.g.][]{jonsson_sunrise:_2006, baes_efficient_2011,
  robitaille_hyperion:_2011, steinacker_three-dimensional_2013}, that we did
not include any modelling of the dust re-emission or of the change in
dust temperature (and properties).

We used the radiation-hydrodynamics SPHINX simulations
\citep{rosdahl_sphinx_2018}, the S10\_512\_BINARY run in
practise. We extracted from the output at $z=6$ a spherical data
domain centred onto the most massive halo and with a radius equal to
$1.1 R_{vir}$. This galaxy has a stellar mass of
$\sim 1.5 \times 10^9 M_\odot$ and a star formation rate of
$\sim 5 M_\odot/$yr. For the \texttt{gas\_composition} we used the
default values for the dust composition (SMC model, $f_{ion} = 0.01$,
$Z_{ref} = 0.005$).

For the emission of photons, we used the \texttt{PhotonsFromStars}
spatial sampling and the tabulated continuum from stellar populations
for the spectral sampling. To be consistent with the ionising
radiative transfer made during the course of the simulation, we used
the stellar library by \citet{eldridge_binary_2017}.  The emitted
rest-frame spectrum was decomposed into different parts corresponding to
the different photometric filters of the JWST/NIRCam, namely F115W,
F150W, F200W, F277W, F356W, and F444W. Each wavelength range was
sampled with ten million photon packets. We propagated only
  stellar continuum photons, not the contribution of the nebular
emission lines from \ion{H}{ii} regions.

We computed the dust continuum RT using for each filter a constant
value for the albedo $a_{dust}$ and the $g$ parameter according to
\citet[][Table 6]{li_infrared_2001}. We again used the
  peeling algorithm to collect flux into a 3D cube, which is eventually
integrated along the spectral dimension using the throughputs of
various NIRCam
filters\footnote{\texttt{https://jwst-docs.stsci.edu/display/JTI/NIRCam+Filters}}
to obtain images. We then dimmed the image using the luminosity distance
of the galaxy (at $z=6$).

In Fig.~\ref{fig:rgb}, we show a pseudo-colour image combining F150W,
F277W, and F444W images at a very high resolution. The effect of
  dust can be seen with the dust lane in the edge-on view (bottom
  panel). We also show mock images in the F150W and F444W filters
with a noise distribution at a level corresponding to a time exposure
of 1 Ms. The two rows show two different projections of the same
object. The bottom row is the same projection as in the maps of
\citet[][Fig. 4]{rosdahl_sphinx_2018}.

\section{Summary and conclusions}
\label{sec:summary}

In this paper, we have presented a new public 3D Monte Carlo code called
\ras{} to compute radiative transfer of resonant lines in simulations
of astrophysical objects. \ras{} is written in modern Fortran. The main features of \ras{} are the
following:
\begin{enumerate}
\item \ras{} computes radiative transport of resonant-line photons
  through complex mixes of species (e.g. \ion{H}{i}, \ion{Si}{ii},
  \ion{Mg}{ii}) and dust. Although it is designed to 
  accurately describe resonant scattering, \ras{} may also be used to propagate
  photons at any wavelength (e.g. stellar continuum or fluorescent
  lines).

\item \ras{} performs RT on an adaptive mesh with an octree structure
  very similar to that used and produced by \textsc{ramses}. With very
  little effort, \ras{} can be extended to use any (irregular) mesh
  structure.

\item \ras{} can be easily used to perform RT experiments through
  idealised gas distributions (e.g. expanding shells, discs)
  instead of simulations.

\item \ras{} includes tools which allow the user to sample different
  sources of radiation. In the current distribution, these include 
  a code to spawn photon packets from star particles, taking into
  account their ages and metallicites and using various SED libraries,
  and  a code to spawn \lya{} photons emitted by the gas, taking
  into account both recombination and collisional
  contributions. \ras{} also features a simple code to produce ad hoc
  sources which may be used for tests or idealised models.

\item The default outputs of \ras{} are lists of photon packets
  collected when they escape the computational domain. These are
  typically analysed through a python class that is provided in the
  distribution.

\item \ras{} also features a standard peeling algorithm which allows
  the user to construct mock observations in the form of spectra,
  images, or datacubes on-the-fly.

\item The modularity of \ras{} makes it very easy to implement new
  transitions and to compute RT through new mixtures of scatterers and
  dust. It also makes \ras{} a powerful toolbox to analyse AMR
  simulations in general. For example, it has been used to compute
  escape fractions from galaxies of the SPHINX simulations
  \citep{rosdahl_sphinx_2018} by casting hundreds of rays from each
  star particle in the simulation and integrating the optical depths
  to ionising radiation along these rays up to the virial radii of
  host halos.

\item \ras{} is parallelised using MPI and shows perfect scaling at
  least up to a thousand cores. It also features domain decomposition,
  which  reduces its memory footprint to arbitrarily low
  values. These features make it usable to process simulations of
  arbitrarily large sizes on large supercomputers.

\item \ras{} has been fully tested against RT problems with analytic
  solutions and against various test cases proposed in the
  literature. In all cases, the agreement between \ras{} and published
  results is very good.

\item The \ras{} code is publicly available\footnote{\texttt{http://rascas.univ-lyon1.fr}}. Future developments
  of the code will be released and documented at the same URL. \ras{}
  has been designed to be both easy to use and easy to develop, and we
  hope to share future developments with a growing community.

\end{enumerate}

\begin{acknowledgements}
  The authors are happy to acknowledge useful and stimulating
  discussions with H. Katz, C. Scarlata, J. Prochaska, A. Henry,
  A. Smith, M. Haehnelt, L. Barnes. We further kindly thank
  B. S\'emelin for providing us with parts of his own \lya{} RT
  code. We thank J. Rosdahl and V. Mauerhofer for their valuable
  contributions to improving the code. We acknowledge development of
  some of the metal-line modules from intern students J. Dumoulin,
  C. Dubois, A. Collard.  We also acknowledge Franz Schreier for
  making publicly available his python code for comparing
  approximations of the Voigt function. We thank the anonymous referee
  for her/his careful and constructive report.
  \\
  JB acknowledge support from the ORAGE project from the Agence
  Nationale de la Recherche under grant ANR-14-CE33-0016-03.  TG is
  grateful to the LABEX Lyon Institute of Origins (ANR-10-LABX-0066)
  of the Univesit\'e de Lyon for its financial support within the
  program “Investissements d’Avenir” (ANR-11-IDEX-0007) of the French
  government operated by the National Research Agency (ANR).  AV
  acknowledges support from the MHV-SNF grants PP00P2\_1176808,
  PMPDP2\_175707, and from the European Research Council under grant
  agreement ERC-stg-757258 (TRIPLE).  TK was supported in part by the
  National Research Foundation of Korea (No. 2017R1A5A1070354 and
  No. 2018036146) and in part by the Yonsei University Future-leading
  Research Initiative (RMS2-2018-22-0183).  MT acknowledges funding
  from the European Research Council under the European Community's
  Seventh Framework Programme (FP7/2007-2013 Grant Agreement
  no. 614199, project 'BLACK').
  \\
  The development and tests of \ras{} were greatly facilitated by our
  access to computing resources at the Common Computing Facility (CCF)
  of the LABEX Lyon Institute of Origins (ANR-10-LABX-0066), and at
  the CC-IN2P3 Computing Centre (Lyon/Villeurbanne - France), a
  partnership between CNRS/IN2P3 and CEA/DSM/Irfu.

\end{acknowledgements}

%
\bibliographystyle{aa} 
\bibliography{bibtex_rascas_paper} 
%

\begin{appendix} 

\section{Atomic data}

In Table~\ref{table:1}, we provide atomic data for a selected sample
of species and transitions implemented in \textsc{rascas}.

\begin{table*}
  \caption{Atomic data for the species and transitions implemented in
    \textsc{rascas}. Each group (between two horizontal lines) shows one
    absorption line and the decay channel(s) (resonant and fluorescent
    if any). Data taken from the NIST database
    \citep[][\texttt{https://www.nist.gov/}]{NIST_ASD}.}
\label{table:1}      
\centering                          
\begin{tabular}{c l c l l l l}        
  \hline\hline                 
  Species & Transition nickname & Vac. Wavelength & $A_{ul}$ & $f_{lu}$ & Lower Level ($l$) & Upper Level ($u$) \\   
    &   & ($\AA$) & (s$^{-1}$) & & \\
\hline                        
\ion{H}  & Ly$\alpha$ & $1215.67$ &$ 6.265 \times 10^8$ & $ 0.416$ &  $1s$ & $2p$ \\
\hline                                   
\ion{D}  & D Ly$\alpha$ & $1215.34$ &$ 6.265 \times 10^8$ & $ 0.416$ &  $1s$ & $2p$ \\
\hline                                   
\ion{Si} & \ion{Si}{ii} $\lambda 1190$             & $1190.42$ & $6.53 \times 10^8$ & $0.277$  & $3s^23p$~~\element[\mathrm{o}][2]{P}~~1/2 & $3s3p^2$~~\element[][2]{P}~~3/2\\
\ion{Si} & \ion{Si}{ii}$^{\star}$ $\lambda 1194$    & $1194.50$ & $3.45 \times 10^9$ &  \_      & $3s^23p$~~\element[\mathrm{o}][2]{P}~~3/2 & $3s3p^2$~~\element[][2]{P}~~3/2 \\
\hline                                   
\ion{Si} & \ion{Si}{II} $\lambda 1193$             & $1193.28$ & $2.69 \times 10^9$ & $0.575$  & $3s^23p$~~\element[\mathrm{o}][2]{P}~~1/2 & $3s3p^2$~~\element[][2]{P}~~1/2\\
\ion{Si} & \ion{Si}{II}$^{\star}$ $\lambda 1197$    & $1197.39$ & $1.40 \times 10^9$ &  \_      & $3s^23p$~~\element[\mathrm{o}][2]{P}~~3/2 & $3s3p^2$~~\element[][2]{P}~~1/2 \\
\hline                                   
\ion{Si} & \ion{Si}{II} $\lambda 1260$             & $1260.42$ & $2.57 \times 10^9$ & $1.22$   & $3s^23p$~~\element[\mathrm{o}][2]{P}~~1/2 & $3s^23d$~~\element[][2]{D}~~3/2\\
\ion{Si} & \ion{Si}{II}$^{\star}$ $\lambda 1265$    & $1265.02$ & $4.73 \times 10^8$ &  \_      & $3s^23p$~~\element[\mathrm{o}][2]{P}~~3/2 & $3s^23d$~~\element[][2]{D}~~3/2 \\
\hline                                   
\ion{Mg} & \ion{Mg}{II} $\lambda 2796$             & $2796.35$ & $2.60 \times 10^8$ & $0.608$  & $2p^63s$~~\element[][2]{S}~~1/2 & $2p^63p$~~\element[\mathrm{o}][2]{P}~~3/2\\
\hline                                   
\ion{Mg} & \ion{Mg}{II} $\lambda 2804$             & $2803.53$ & $2.57 \times 10^8$ & $0.303$  & $2p^63s$~~\element[][2]{S}~~1/2 & $2p^63p$~~\element[\mathrm{o}][2]{P}~~1/2\\
\hline                                   

\ion{Fe} & \ion{Fe}{II} $\lambda 2250$             & $2249.88$ & $3.00 \times 10^6$ &  $0.00182$ & $3d^64s$~~\element[][6]{D}~~9/2 & $3d^64p$~~\element[\mathrm{o}][4]{D}~~7/2 \\
\ion{Fe} & \ion{Fe}{II}$^{\star}$ $\lambda 2270$    & $2269.52$ & $4.00 \times 10^5$ &  \_        & $3d^64s$~~\element[][6]{D}~~7/2 & $3d^64p$~~\element[\mathrm{o}][4]{D}~~7/2 \\
\hline
\ion{Fe} & \ion{Fe}{II} $\lambda 2261$             & $2260.78$ & $3.18 \times 10^6$ &  $0.00244$ & $3d^64s$~~\element[][6]{D}~~9/2 & $3d^64p$~~\element[\mathrm{o}][4]{F}~~9/2 \\
\ion{Fe} & \ion{Fe}{II}$^{\star}$ $\lambda 2281$    & $2280.62$ & $4.49 \times 10^6$ &  \_        & $3d^64s$~~\element[][6]{D}~~7/2 & $3d^64p$~~\element[\mathrm{o}][4]{F}~~9/2 \\
\hline
\ion{Fe} & \ion{Fe}{II} $\lambda 2344$             & $2344.21$ & $1.73 \times 10^8$ & $0.114$  & $3d^64s$~~\element[][6]{D}~~9/2 & $3d^64p$~~\element[\mathrm{o}][6]{P}~~7/2 \\
\ion{Fe} & \ion{Fe}{II}$^{\star}$ $\lambda 2366$    & $2365.55$ & $5.90 \times 10^7$ &  \_      & $3d^64s$~~\element[][6]{D}~~7/2 & $3d^64p$~~\element[\mathrm{o}][6]{P}~~7/2 \\
\ion{Fe} & \ion{Fe}{II}$^{\star}$ $\lambda 2381$    & $2381.49$ & $3.10 \times 10^7$ &  \_      & $3d^64s$~~\element[][6]{D}~~5/2 & $3d^64p$~~\element[\mathrm{o}][6]{P}~~7/2 \\ 
\hline
\ion{Fe} & \ion{Fe}{II} $\lambda 2374$             & $2374.46$ & $4.25 \times 10^7$ & $0.0359$ & $3d^64s$~~\element[][6]{D}~~9/2 & $3d^64p$~~\element[\mathrm{o}][6]{F}~~9/2 \\
\ion{Fe} & \ion{Fe}{II}$^{\star}$ $\lambda 2396$    & $2396.35$ & $2.59 \times 10^8$ &  $\_$    & $3d^64$s~~\element[][6]{D}~~7/2 & $3d^64p$~~\element[\mathrm{o}][6]{F}~~9/2 \\
\hline                                   
\ion{Fe} &  \ion{Fe}{II} $\lambda 2383$            & $2382.76$ & $3.13 \times 10^8$ &  $0.32$  & $3d^64s$~~\element[][6]{D}~~9/2 & $3d^64p$~~\element[\mathrm{o}][6]{F}~~11/2 \\
\hline                                   
\ion{Fe} & \ion{Fe}{II} $\lambda 2587$             & $2586.65$ & $8.94 \times 10^7$ & $0.0717$ & $3d^64s$~~\element[][6]{D}~~9/2 & $3d^64p$~~\element[\mathrm{o}][6]{D}~~7/2 \\
\ion{Fe} & \ion{Fe}{II}$^{\star}$ $\lambda 2612$    & $2612.65$ & $1.20 \times 10^8$ &  \_      & $3d^64s$~~\element[][6]{D}~~7/2 & $3d^64p$~~\element[\mathrm{o}][6]{D}~~7/2 \\
\ion{Fe} & \ion{Fe}{II}$^{\star}$ $\lambda 2632$    & $2632.11$ & $6.29 \times 10^7$ &  \_      & $3d^64s$~~\element[][6]{D}~~5/2 & $3d^64p$~~\element[\mathrm{o}][6]{D}~~7/2 \\
\hline                                   
\ion{Fe} & \ion{Fe}{II} $\lambda 2600$             & $2600.17$ & $2.35 \times 10^8$ &  $0.239$ & $3d^64s$~~\element[][6]{D}~~9/2 & $3d^64p$~~\element[\mathrm{o}][6]{D}~~9/2 \\
\ion{Fe} & \ion{Fe}{II}$^{\star}$ $\lambda 2626$    & $2626.45$ & $3.52 \times 10^7$ &  \_      & $3d^64s$~~\element[][6]{D}~~7/2 & $3d^64p$~~\element[\mathrm{o}][6]{D}~~9/2 \\
\hline                                   
\end{tabular}
\end{table*}

\section{Additional tests}

\begin{figure}
  \centering
  \includegraphics[width=8.9cm]{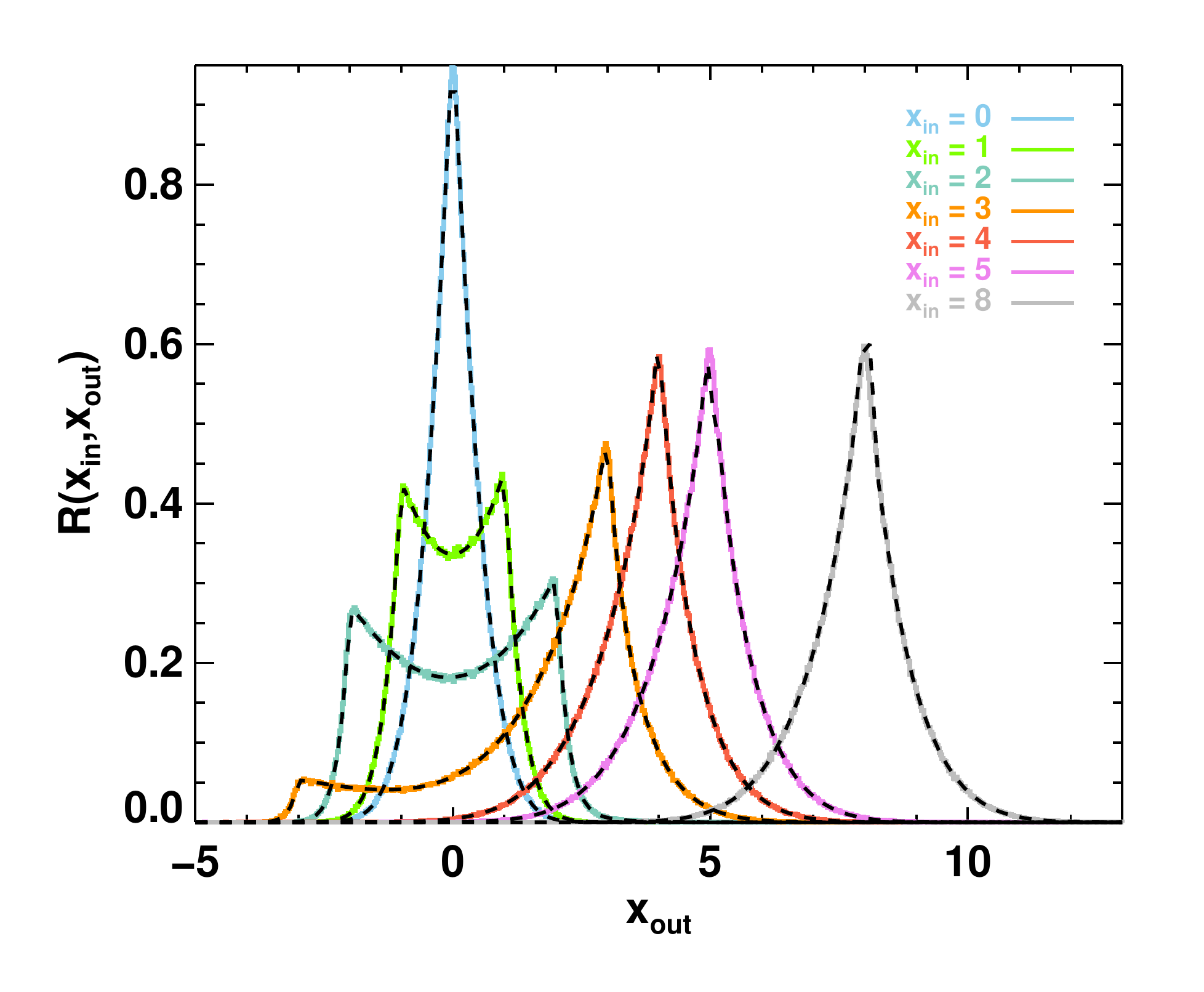}
   \caption{Frequency redistribution, $R(x_{\rm in},x_{\rm out})$, in
     Doppler units $x_{\rm out}$ of Lyman$-\alpha$ photons emitted at
     $x_{\rm in}$ after one single scattering on a hydrogen atom assuming
     Rayleigh angular redistribution. The solid coloured
     curves are the results from \textsc{rascas} assuming
     $x_{\rm in}=0, 1, 2, 3, 4, 5$, and 8, while the dashed black lines
     are the analytic solutions of \citet{hummer_non-coherent_1962}.}
\label{FreqRedist_rayleigh_D}
\end{figure}

In Sect.~\ref{test_cases}, we  present a series of tests based
on the comparison of \textsc{rascas} with analytic solutions and other
MC codes. In Fig.~\ref{indiv_scatt}, we show the frequency
redistribution of Lyman$-\alpha$ photons after one scattering on a
hydrogen atom assuming isotropic angular redistribution. Here we
repeat the same experiment (Fig.~\ref{FreqRedist_rayleigh_D}), but we
now check the frequency redistribution function for Rayleigh
scattering on hydrogen atoms. We find perfect agreement with the
expected solution of \citet{hummer_non-coherent_1962}.

\begin{figure}
  \centering
  \includegraphics[width=7.8cm]{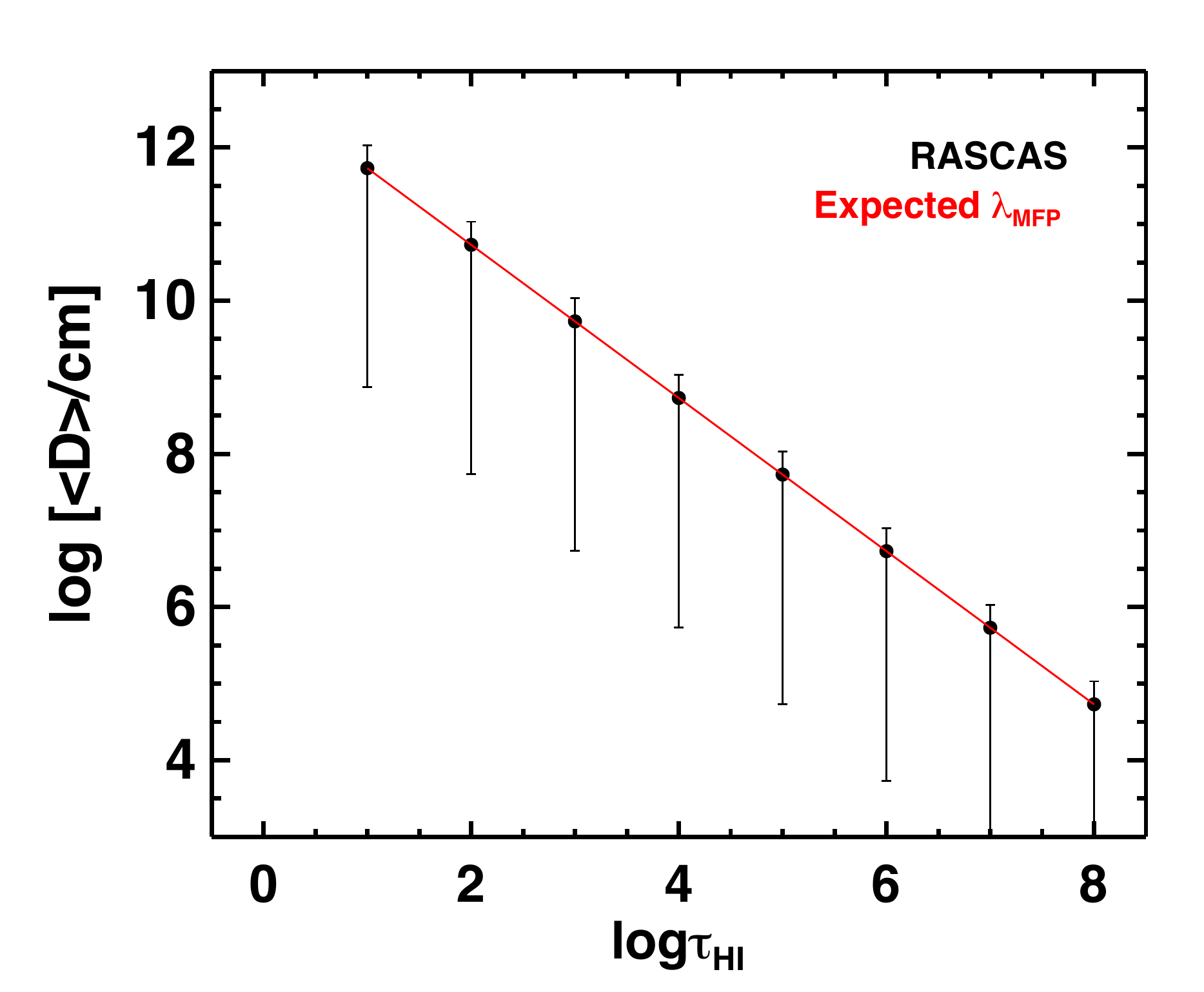}
  \caption{Distance travelled by Lyman$-\alpha$ photons emitted at
    $x_{\rm in}=0$ before their first scattering as a function of
    \ion{H}{I} opacity. The mean travelled distance (filled circles)
    is in excellent agreement with the expected mean free path
    ($\lambda_{\rm MFP} = (n_{\rm \hi} \sigma_{0})^{-1}$; red line) at
    all opacities. Error bars correspond to the $1\sigma$ standard
    deviation.}
\label{mfp}
\end{figure}

Figure~\ref{mfp} compares the mean distance travelled by $10^6$
Lyman$-\alpha$ photons emitted at $x_{\rm in}=0$ as a function of
line-centre opacity $\tau_{\rm \hi}$ (in black) with the mean free
path defined as $\lambda_{\rm MFP} = (n_{\rm \hi} \sigma_{0})^{-1}$
(red curve), where $n_{\rm \hi}$ and $\sigma_{0}$ are the \ion{H}{i} density
of the uniform medium and the Lyman$-\alpha$ cross section at line
centre for $T=100$~K. Again, we confirm that \textsc{rascas} recovers
well the expected value from the optically thin ($\tau_{\rm \hi}=10$)
to extremely thick regime ($\tau_{\rm \hi}=10^8$).

\begin{figure}
  \centering
  \includegraphics[width=9.8cm]{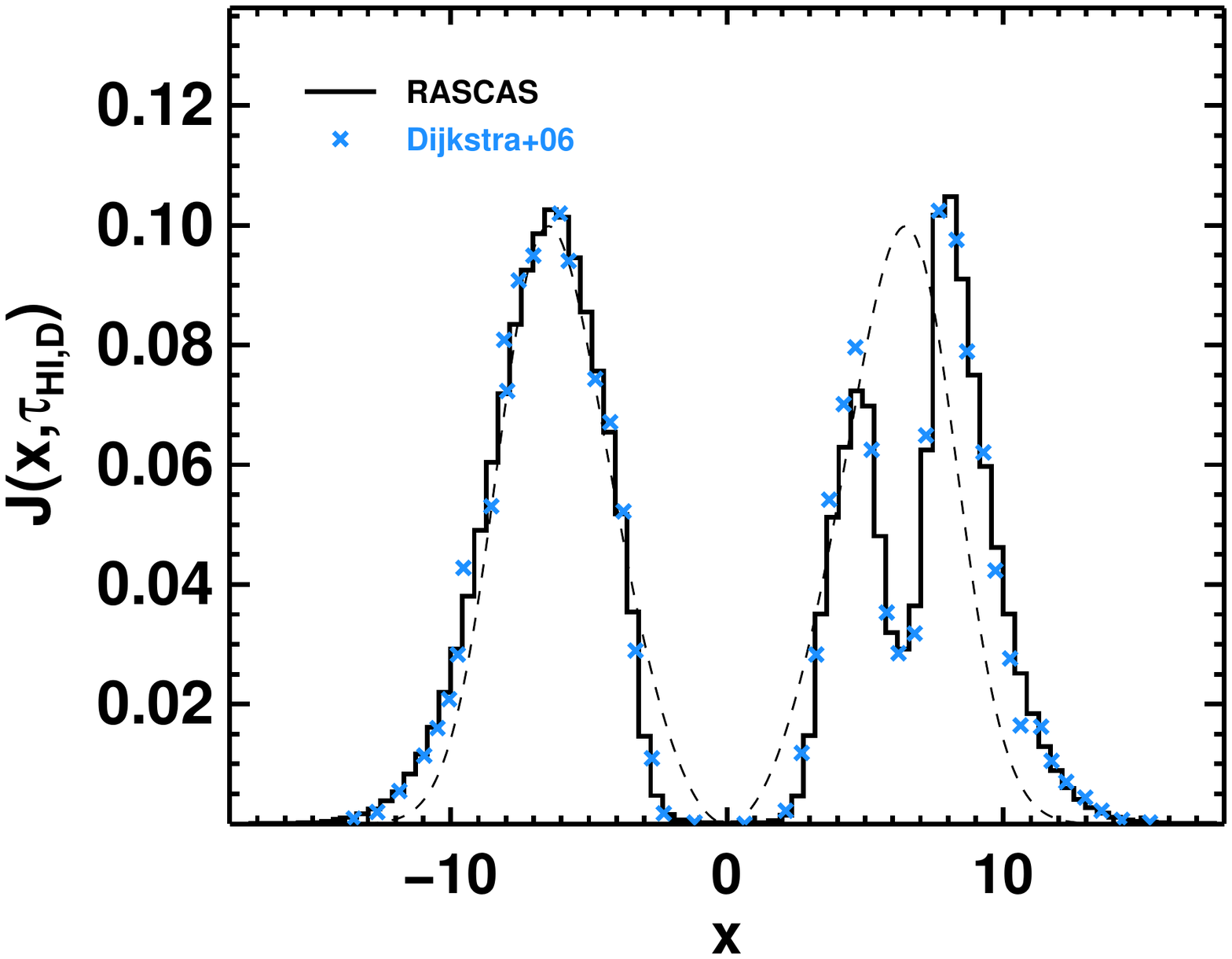}
  \vspace{-3.4cm}
  \caption{Emergent spectrum of Lyman$-\alpha$ photons
    ($x_{\rm in}=0$) emitted at the centre of a uniform static sphere
    of \ion{H}{I} mixed with neutral deuterium D. The model parameters
    are similar to those used in Figure 3 of \citet{dijkstra_ly_2006}:
    $T=10^4$~K, $N_{\ion{H}{I}}=2 \times 10^{19}$ cm$^{-2}$, and
    $D/H = 3 \times 10^{-5}$. We recover the same
    absorption feature as \citet{dijkstra_ly_2006} on the blue side of
    the Lyman$-\alpha$ line ($x \approx 6.3$), which is due to the
    presence of deuterium, compared to the case for pure \hi\ (dashed line).}
  \label{sphere_deuterium_effect}
\end{figure}

\textsc{rascas} can also be used to follow radiative transfer through
a medium composed of several elements simultaneously. This can be
particularly useful if different elements exhibit atomic transitions
with comparable energy levels, for example  in the case of the \lya{} line of
atomic hydrogen ($\lambda_{\rm Ly\alpha} = 1215.67$ \AA) and deuterium
($\lambda_{\rm D, Ly\alpha} = 1215.34$ \AA). In
Fig.~\ref{sphere_deuterium_effect} we show the spectrum in $x$ units
(normalised to the hydrogen line centre) emerging from a static sphere
of \ion{H}{I} mixed with deuterium, assuming an abundance of
$\textrm{D}/\textrm{H} = 3 \times 10^{-5}$ computed with
\textsc{rascas} (solid black curve). It  can be seen that the presence of
deuterium affects the line profile expected for pure \hi\ (i.e. a
symmetric double peak; dashed curve in
Fig.~\ref{sphere_deuterium_effect}) by producing a sharp absorption
feature in the blue peak of the spectrum at $x \approx 6.3$, which
corresponds to the \lya{} resonance of deuterium.  For comparison, we
overplot the simulated spectrum of \citet{dijkstra_ly_2006} for the
same experiment (blue crosses), and find a very good agreement between
the two.

\end{appendix}

\end{document}